\documentclass[a4paper, amsfonts, amssymb, amsmath, reprint, showkeys, nofootinbib, twoside]{revtex4-1}
\usepackage[english]{babel}
\usepackage[utf8]{inputenc}
\usepackage[colorinlistoftodos, color=green!40, prependcaption]{todonotes}

\usepackage{algorithm}
\usepackage{algpseudocode}
\algnewcommand\algorithmicswitch{\textbf{switch}}
\algnewcommand\algorithmiccase{\textbf{case}}
\algnewcommand\algorithmicassert{\texttt{assert}}
\algnewcommand\Assert[1]{\State \algorithmicassert(#1)}%
\algdef{SE}[SWITCH]{Switch}{EndSwitch}[1]{\algorithmicswitch\ #1\ \algorithmicdo}{\algorithmicend\ \algorithmicswitch}%
\algdef{SE}[CASE]{Case}{EndCase}[1]{\algorithmiccase\ #1}{\algorithmicend\ \algorithmiccase}%
\algtext*{EndSwitch}%
\algtext*{EndCase}%

\usepackage{amsthm}
\usepackage{mathtools}
\usepackage{physics}
\usepackage{xcolor}
\usepackage{graphicx}
\usepackage[left=23mm,right=13mm,top=35mm,columnsep=15pt]{geometry} 
\usepackage{adjustbox}
\usepackage{placeins}
\usepackage[T1]{fontenc}
\usepackage{lipsum}
\usepackage{csquotes}

\usepackage[pdftex, pdftitle={Article}, pdfauthor={Author}]{hyperref} 
\begin{document}
\title{Unveiling the entropic role of hydration water in SOD1 partitioning within FUS condensate.}

\author{Luis Enrique Coronas}
\affiliation{ Departament de F\'isica de la Mat\`eria Condensada, Facutat de F\'isica, University of Barcelona, Mart\'i i Franqu\`es 1, Barcelona 08028, Spain}
\affiliation{Institut de Nanoci\`encia i Nanotecnologia (IN2UB), Universitat de Barcelona, Mart\'i i Franqu\`es 1, Barcelona 08028, Spain}
\affiliation{Université Paris Cité, CNRS, Laboratoire de Biochimie Théorique, 13 rue Pierre et Marie Curie, 75005 Paris, France}

\author{Stepan Timr}
\affiliation{Université Paris Cité, CNRS, Laboratoire de Biochimie Théorique, 13 rue Pierre et Marie Curie, 75005 Paris, France}
\affiliation{J. Heyrovsky Institute of Physical Chemistry, Czech Academy of Sciences, Dolejskova 2155/3, 182 00 Prague, Czechia}

\author{Fabio Sterpone}
\affiliation{Université Paris Cité, CNRS, Laboratoire de Biochimie Théorique, 13 rue Pierre et Marie Curie, 75005 Paris, France}

\author{Giancarlo Franzese}
\affiliation{ Departament de F\'isica de la Mat\`eria Condensada, Facutat de F\'isica, University of Barcelona, Mart\'i i Franqu\`es 1, Barcelona 08028, Spain}
\affiliation{Institut de Nanoci\`encia i Nanotecnologia (IN2UB), Universitat de Barcelona, Mart\'i i Franqu\`es 1, Barcelona 08028, Spain}

\email{gfranzese@ub.edu}


\begin{abstract}
Biological processes like the sequestration of Superoxide Dismutase 1 (SOD1) into biomolecular condensates such as FUS and stress granules are essential to understanding disease mechanisms, including amyotrophic lateral sclerosis (ALS). Our study demonstrates that the hydration environment is crucial in these processes. Using the advanced CVF water model, which captures hydrogen-bond networks at the molecular level, we show how water greatly impacts SOD1’s behavior, residency times, and transition rates between different associative states. Importantly, when water is included to hydrate an implicit solvent model (OPEP), we gain a new perspective on the free energy landscape of the system, leading to a conclusion that clarifies that suggested by OPEP alone. While the OPEP model indicated that Bovine Serum Albumin (BSA) crowders reduce SOD1’s partition coefficient (PC) mainly due to nonspecific interactions with BSA, our enhanced explicit-water approach reveals that the hydration entropy behavior in BSA drives the observed decrease in PC. This highlights that explicitly modeling water is essential for accurately understanding protein-crowder interactions and their biological relevance, emphasizing water’s role in cellular phase separation and disease-related processes.
\end{abstract}

\maketitle

\section{Introduction}
\label{OPEP:Introduction_sod1}

Biomolecular condensates organize the cellular fluid and play a vital role in cell functions like ribosome biogenesis or DNA damage response \cite{Banani2017}. These membraneless organelles (MLOs) represent the droplet phase resulting from protein liquid-liquid phase separation (LLPS) \cite{Brangwynne2009, vanLeeuwen2019, Andre2020}. Experimental studies reveal that many MLOs share liquid-like properties, including fusion \cite{Brangwynne2009}, dripping \cite{Brangwynne2009}, high viscosity \cite{Elbaum-Garfinkle2015}, or wetting \cite{Kusumaatmaja2021}. Other research has examined physico-chemical factors influencing droplet stability, such as temperature 
\cite{Wang2012}, pH \cite{Franzmann2018}, or ionic strength \cite{Thompson2016}, raising growing interest in a multidisciplinary community \cite{Alberti:2025aa}.

Stress granules (SGs) are MLOs formed under cell stress conditions, such as starvation \cite{vanLeeuwen2019} or heat stress (HS) \cite{Riback2017}. Proteins are known to unfold either completely or partially when heated, and misfolded proteins can aggregate into toxic assemblies \cite{Zhao2019}. SGs serve as a cytoprotective mechanism, sequestering misfolded proteins or aggregates \cite{Gallardo2021, Vabulas2010, Wood2018, Ganesan2008} and thus preventing disease development. For example, Superoxide Dismutase 1 (SOD1) proteins, which are linked to the progression of amyotrophic lateral sclerosis (ALS) \cite{Wright2019, Nguyen2021}, are sequestered {\it in vivo} into SGs and {\it in vitro} into Fused in Sarcoma (FUS) condensates \cite{Samanta2021} upon heating.

Low complexity domains (LCD) of intrinsically disordered proteins (IDPs), such as FUS \cite{Burke2015, Thompson2018, Kamagata2022}, heterogeneous nuclear ribonucleoprotein A1 (hnRNPA1) \cite{Molliex2015, Tsoi2021}, or DEAD-box helicase protein LAF-1 \cite{Elbaum-Garfinkle2015}, commonly drive LLPS \cite{Elbaum-Garfinkle2015, Quiroz2015, Dignon2018}. Although simple {\it in vitro} experiments show that homotypic FUS interactions are sufficient to induce phase separation \cite{Burke2015}, the cellular environment is a complex, crowded, multi-component mixture of biomolecules. Crowded environments typically promote LLPS through excluded volume effects, co-condensation of crowding agents and phase-separating biomolecules, or by inducing segregative phase separation \cite{Andre2020}. In {\it in vitro} experiments, synthetic polymer molecules such as Ficoll 70 or polyethylene glycol (PEG) are used to mimic cellular crowding agents, or crowders (CWDs) \cite{Rivas2016}. However, these molecules are relatively inert with respect to the phase-separating protein, ignoring the chemical specificity of the protein-CWD interactions. Bovine Serum Albumin (BSA) provides a step toward a more realistic mimicking of cytoplasm crowding in {\it in vitro} experiments \cite{Gnutt2019, Samanta2021, Kanaan2020}, as it is highly water-soluble and introduces, to some extent, protein-protein interactions similar to those in the cellular environment \cite{Kuznetsova2015}. 

In this work, we study the hydration effects on SOD1 sequestration in FUS condensates and in BSA crowded media. 
In line with previous work \cite{Samanta2021}, we focus on a loop-truncated monomeric variant of SOD1 \cite{Danielsson:2011aa}, which facilitates experimental measurements because of its reduced aggregation propensity and higher thermal stability.
Fully atomistic simulations have investigated the driving forces of biomolecular phase separation related to solvation thermodynamics, but their high computational cost often restricts them to simulating only a few protein molecules \cite{Dignon:2018aa} or individual amino acid side chains \cite{doi:10.1073/pnas.2425422122}.
We consider simulation results from a model, which implicitly accounts for the water contributions \cite{Samanta2021}, and hydrate the system to gain a new perspective that highlights the role of water. In their original study, Samanta et al. conducted {\it in vitro} experiments and simulations to compare the sequestration of SOD1 into various CWDs, motivated by two primary objectives. First, they sought to explain the lower partition coefficient (PC) of SOD1 in FUS condensates when globular protein BSA is used as a CWD, as opposed to synthetic polymer Ficoll 70. Second, the comparison is significant due to the distinct sequence-specific interactions, crowding effects, excluded volume influences, and solvent compositions in the different environments. For the globular BSA, most of the volume is occupied by a reservoir of bulk water, whereas FUS, an IDP, is homogeneously distributed throughout the entire system volume, resulting in larger excluded volume effects experienced by SOD1.

Despite these differences, simulations—using hydrodynamic effects incorporated via the Lattice Boltzmann Molecular Dynamics (LBMD) method \cite{10.1063/1.480156, 10.5555/162737.162773}—of the OPEP model \cite{SterponeJCTC2015, SterponeJPCB2018}—an implicit-solvent, coarse-grained protein model that preserves sequence specificity and assumes short-range interactions \cite{SterponeReview2014}—did not reveal significant differences in the SOD1-CWD interaction energies. 
However, the experiments showed a slightly stronger preference for partitioning into BSA compared with FUS \cite{Samanta2021}. Samanta et~al. concluded that the minor PC of SOD1 in FUS condensates within BSA-crowded medium, compared to Ficoll 70, is explained by the finding that SOD1’s interaction energy with BSA is at least comparable in magnitude to its interaction energy with FUS.

Water, especially hydrogen bonds (HBs) at the hydration shell, influences the free-energy landscape of biomolecular systems \cite{Chaplin, Ball2017}. However, including water molecules explicitly in simulations is often too computationally demanding, especially in biological environments where millions of water molecules interact with biomolecules. To address this issue, implicit-solvent models are employed; these models integrate over the system's solvation degrees of freedom (DoF), capturing water effects through effective potentials. This approach hides the solvation coordinates, interpreting all entropy and enthalpy contributions as part of the internal energy of protein-protein interactions. By integrating out the water DoF, implicit models omit explicit water molecules but still account for their influence in potentials and free energy calculations. 

Here, we hypothesize that including explicit solvent interactions is essential to elucidate the different behaviors of SOD1 sequestered in FUS and BSA environments.
We apply the CVF model for water, which accounts for HB interactions at the molecular level, including cooperativity, while coarse-graining the positions of the molecules \cite{coronas2024phase, CoronasJMolLiq2025, coronas2025algorithms}. Thanks to a parametrization based on {\it ab initio} calculations and experimental data, the model quantitatively reproduces the experimental water thermodynamic behavior within a range of 60 degrees around ambient temperature and from atmospheric pressure up to 50 MPa \cite{CoronasJMolLiq2025}. The equation of state, extended to deep supercooled conditions, aligns with atomistic and polarizable models \cite{coronas2024phase}.

Following the method described in Sec.~\ref{couple_cvf_opep}, we hydrate configurations generated by the implicit-solvent OPEP model \cite{SterponeReview2014, Samanta2021} (Sec.~\ref{opep}) with CVF water (Sec.~\ref{cvf}). The hydration process does not alter the fundamental physics, particularly the free energy, because the original OPEP results already implicitly account for water effects through effective potentials. However, by explicitly coupling OPEP configurations with water molecules, we can directly estimate the solvent's contribution to the system's free energy. This approach allows us to project the free energy onto water-dependent coordinates, uncovering features of the free energy landscape that are not apparent in implicit solvent models. Notably, we identify three characteristic associative states of crowded SOD1 proteins, along with their corresponding free-energy barriers and transition rates. This new perspective highlights the role of water entropy in driving SOD1 sequestration.

\section{Model and methods}
\label{model_methods}

\subsection{Implicit solvent OPEP simulations}
\label{opep}

We describe proteins using the OPEP coarse-grained model, originally developed to study peptide and protein folding without specific biases and aggregation \cite{SterponeReview2014, Chebaro2012, SterponeJCTC2013}. Early versions of the model represented the backbone atomistically and the side chain as a single bead. However, when the main conformation remains stable over time, it is preferable to use an elastic network model based on the OPEP force field for intramolecular non-bonded interactions, with a simplified resolution of two beads per residue: side chain and C$\alpha$. Alanine, proline, and glycine are exceptions, as they are only represented with the C$\alpha$. The elastic network's purpose is to maintain the native structure of folded conformations. We apply the elastic network version of OPEP to SOD1, as shown in Ref. \cite{Timr2021}, and to BSA proteins. The force field includes bonded 
and non-bonded interactions. In this work, we adopt OPEPv4 for non-bonded interactions \cite{SterponeReview2014}, with a refinement for oppositely charged residue pairs \cite{Timr2021}. This version is essentially identical to OPEPv7, except for cysteine interactions not present in SOD1, created to replicate the diffusivity of globular proteins in crowded protein solutions \cite{Timr2023}. 

For the low-complexity domain (LCD) of FUS, we use a coarse-grained model inspired by the single-bead per residue approach for intrinsically disordered proteins \cite{Dignon2018}. Neighboring residues are connected through a harmonic spring, and non-bonded interactions are set as described for the refined OPEPv4 above. A similar flexible-chain version of OPEPv7 predicts LLPS and has been employed to study rheological properties of biomolecular condensates under shear \cite{Coronas2024b}.

To incorporate hydrodynamic effects, we couple the Lattice Boltzmann (LB) grid representation of the fluid with Langevin molecular dynamics (MD). LBMD simulations of the implicit solvent OPEP model \cite{SterponeReview2014, SterponeJCTC2015, SterponeJPCB2018} have been performed to analyze SOD1 sequestration into BSA \cite{Gnutt2019, Timr2020}, as well as Chymotrypsin Inhibitor 2 (CI2) sequestration into BSA and Lysozyme \cite{Timr2021} crowded environments. In this work, we consider protein conformations sampled through LBMD simulations of SOD1 in BSA and FUS-LCD highly concentrated solutions \cite{Samanta2021}.

\subsection{The CVF model for water thermodynamic calculations}
\label{cvf}

The CVF water model has recently been developed for thermodynamic calculations in large-scale water systems. The model is quantitatively accurate when compared to experiments for temperatures within a 60-degree range around ambient conditions and for pressures up to 50 MPa \cite{CoronasJMolLiq2025}. Also, its predictions compare well to those of TIP4P/2005 and OPC models \cite{AbascalVegaJCP2005, Izadi2014}, along the atmospheric isobar. Additionally, CVF free energy calculations were extended to supercooled conditions \cite{coronas2024phase}, predicting a liquid-liquid critical point at the thermodynamic limit consistent with finite-size estimates from iAMOEBA \cite{Pathak2016}, TIP4P/Ice \cite{Debenedetti289}, and ML-BOP models \cite{Dhabal_Kumar_Molinero_2024}, as well as an approximation based on a collection of experimental data \cite{Mallamace2024}. Thanks to the implementation of efficient Monte Carlo algorithms parallelized on a GPU, the model can equilibrate water systems containing up to 17 million water molecules that fill a cubic simulation box of $75$~nm edge \cite{coronas2025algorithms}.

The starting hypothesis of the CVF model is that hydrogen bonds (HBs) are the main factor in water's anomalous thermodynamic behavior. To improve the calculation of enthalpy and entropy, the model carefully accounts for HB interactions at the molecular level, while coarse-graining the positions and orientations of molecules through a discrete density field. The CVF model for bulk water is thoroughly described in Refs. \cite{coronas2024phase, coronas2025algorithms, CoronasJMolLiq2025}. Next, we explain the main ideas behind calculating the HB network and discretizing the density. In Sec.~\ref{couple_cvf_opep}, we cover the algorithm for coupling protein configurations with CVF water.

For $N$ water molecules occupying a fluctuating volume $V(T,P)$ at fixed temperature $T$ and pressure $P$, the CVF model partitions $V$ into ${\cal N} \geq N$ cells, each containing up to one water molecule. The volume of each cell is $v_i \geq v_0$, where $i$ is the cell index, and $v_0\equiv r_0^3$. We set $r_0 \equiv 2.9$~\AA, which is the hard-core van der Waals diameter of a water molecule. The ${\cal N}$ cells serve as discrete sites used to calculate the local density influenced by HBs. Specifically, the volume of the cells is decomposed into an isotropic part, $v_{\rm iso}$, and an anisotropic part caused by the presence of HBs. Here, $v_{\rm iso} \equiv r^3$ is the isotropic component and $r$ is the average distance between neighboring water molecules. During the simulation, $r$ is equilibrated according to $T$ and $P$ conditions using a Lennard-Jones potential $U(r)$. This term accounts for all isotropic contributions to the system’s internal energy. 

HBs can form between molecules in neighboring cells $i$ and $j$. As observed in experiments, decreasing the temperature in the liquid phase causes water molecules to form local tetrahedral HBs structures characterized by a larger excluded volume for the central molecule and a lower local density. This structural change does not alter the neighboring distance. Therefore, in the model, the excluded volume of the H-bonded molecules increases, while their relative distance $r\in[r_0,\infty)$ remains unchanged. We observe that for $r>r^{\rm max}$, with $r^{\rm max}\equiv \sqrt[3]{2} r_0 \simeq 3.65$~\AA, water exists in the gas phase \cite{CoronasJMolLiq2025}.
Because the formation of persistent HBs in the gas phase is negligible, the model sets a distance threshold for HB formation at $r\leq r^{\rm max}$. Additionally, Debye-Waller factor estimates \cite{Teixeira1990} and {\it ab initio} simulations that account for nuclear quantum effects \cite{Ceriotti:2013aa} show that HB energy is minimized when $-30^o\leq \widehat{OOH}\leq 30^o$. This indicates that only one-sixth of the possible $\widehat{OOH}$ angles correspond to a bonded state. The model incorporates this reduction in orientational entropy for two H-bonded molecules by limiting their possible orientations to $1/6$ once the HB forms. Finally, experimental and computational studies of liquid water at ambient conditions show that bulk water molecules form up to four tetrahedral HBs in 95\% of cases \citep{DiStasio:2014vu}. Therefore, to simplify, we limit each molecule to a maximum of four HBs at the same time \cite{coronas2025algorithms}. However, since water molecules can quickly switch their HBs between two neighboring molecules within their coordination shell within hundreds of femtoseconds \citep{DamienLaage02102006}, resembling a {\it bifurcated} HB, over-coordinated molecules are also, on average, included in our model.

The formation of a HB increases the global volume $V$ by $v_{\rm HB}$. Therefore, the cell volume is defined as $v_i \equiv v_{\rm iso} + N_{{\rm HB}, i}v_{\rm HB}/2$, where $N_{{\rm HB}, i}$ is the number of HBs formed between cell $i$ and its first neighbors. We divide $v_{\rm HB}$ by $2$ to prevent double counting of the volume.

The non-isotropic HB interaction is further separated into covalent and cooperative contributions. Forming a HB reduces the energy in $J$, which is the characteristic covalent energy of the HB. We set $J\equiv 11$~kJ$/$mol, based on {\it ab initio} energy calculations of isoelectronic molecules at optimal separation \cite{Henry2002}. The cooperative term is $J_\sigma N_\sigma$, representing all many-body contributions to the HB interaction. According to the CVF model, the maximum HB cooperativity occurs at $N_\sigma = 15N$, corresponding to the zero temperature ground state. Based on {\it ab initio} calculations of small water clusters and energy decomposition analysis \cite{Cobar2012}, we find the ratio $J_\sigma/J=0.17$ \cite{CoronasJMolLiq2025}, which means $J_\sigma\equiv 1.87$~kJ$/$mol.

For bulk water, we set $J/(4\epsilon_0)\equiv 0.5$, $J_\sigma/(4\epsilon_0) \equiv 0.08$, and $v_{\rm HB}/v_0 \equiv 0.6$, where $\epsilon_0 \equiv 5.5$~kJ$/$mol is the characteristic energy of the van der Waals interaction. This set of parameters ensures accurate predictions around ambient conditions \cite{CoronasJMolLiq2025}.

The Hamiltonian of the CVF model for bulk water is
\begin{equation}
\begin{aligned}
\label{ham_bulk}
{\cal H}_{\rm bulk} & \equiv {\cal H}_{\rm vdW} + {\cal H}_{\rm HB, covalent} + {\cal H}_{\rm HB, cooperative},
\end{aligned}
\end{equation}
with ${\cal H}_{\rm vdW} \equiv  U(r)$, $ {\cal H}_{\rm HB, covalent}\equiv - JN_{\rm HB}$, and $ {\cal H}_{\rm HB, cooperative}\equiv - J_\sigma N_\sigma$.
The corresponding enthalpy is
\begin{equation}
\begin{aligned}
\label{eq:enthalpy}
H_{\rm bulk} & \equiv {\cal H}_{\rm bulk}  - PV \\
   &\equiv U(r) - JN_{\rm HB} - J_\sigma N_\sigma - P(Nr^3 + N_{\rm HB}v_{\rm HB}).
\end{aligned}
\end{equation}

A detailed formulation of the model, including additional details on the computation of $N_{\rm HB}$ and $N_\sigma$, and its justification can be found in Refs. \cite{coronas2025algorithms, CoronasJMolLiq2025, coronas2024phase}. In particular, we refer to Ref. \cite{CoronasJMolLiq2025} for the setting of the CVF model parameters and rescaling functions that ensure the accuracy of the predictions, and Ref. \cite{coronas2025algorithms} for the implementation of parallel Monte Carlo algorithms, enabling the scalability to large-scale systems.

\subsection{Coupling of CVF water and OPEP protein configurations}
\label{couple_cvf_opep}

We hypothesize that including solvent DoF is essential for accurately analyzing the free energy landscape of biomolecular systems. Here, we test this hypothesis using folded SOD1 proteins sequestered into BSA and FUS crowded environments. Specifically, we adopt protein configurations, representing equilibrium states at ambient conditions, generated by the implicit solvent model OPEP that accounts for the solvent DoF through effective interactions \cite{Samanta2021}. Next, we hydrate these configurations with CVF water and, for each hydrated OPEP system, we generate a MC trajectory equilibrating the water molecules while keeping the proteins fixed in space. The CVF MC dynamics is optimized when the water molecules are partitioned into a compressible (cubic) lattice. Therefore, we develop an algorithm to map arbitrary OPEP configurations into the discretized volume partition used by the CVF model. 

Although the algorithm can be generalized to other solutes such as lipid membranes or nanoparticles, here we focus on residues coarse-grained at a two-bead resolution (side chain and C$_\alpha$), as discussed in Sec.~\ref{opep}. We refer to the mapped OPEP configuration within the discretized grid as "OPEP-dis." Our goal is to keep the OPEP-dis configuration as close as possible to the original OPEP structure. Specifically, we aim to preserve the relative distances and angles among the OPEP beads in the OPEP-dis representation. We note that our spatial resolution is limited by the CVF lattice spacing $r_0=2.9$~\AA.

In the OPEP model, amino acid beads do not have a fixed radius. Instead, they have a repulsion distance that varies depending on the specific bead-bead interactions. For mapping purposes, we assign each amino acid in the OPEP model a specific van der Waals (vdW) volume, $V_{\rm n, vdW}$, where $n$ ranges from 1 to 20, corresponding to each amino acid type \cite{creighton1993}.
To convert an OPEP configuration into its equivalent OPEP-dis configuration, we develop an algorithm described below. In a crowded environment, each cell of the CVF grid can be occupied by more than one amino acid. Therefore, our algorithm uses appropriate probabilities to assign each cell to only one amino acid.
In essence, we associate each OPEP bead—characterized by $V_{\rm n, vdW}$ and centered within a cell of the CVF grid—with the cells whose centers lie within a radius $r_n=\left(3V_{\rm n, vdW}/4\pi\right)^{1/3}$ (Table~\ref{table:CVFExtended_parameters_BiancoPRX}) from the bead's center, assigning higher probability to smaller beads that are closer to the cell as a way of preventing small beads from disappearing. 

The detailed algorithm is the following.
Given an implicit-solvent OPEP configuration, we partition its volume $V$ into ${\cal N}\equiv V/r_0^3$ cubic cells of lateral edge $r_0$. We generate the corresponding OPEP-dis configuration starting from the side-chain beads. The $i$-th cell is 
occupied by the $k$-th side-chain bead with probability
\begin{equation}
\label{eq:GaussianProbability}
p_{ki} \propto \frac{\exp\left( -d_{ki}^2/2 r_k^2\right)}{2\pi r_k},
\end{equation}
where $d_{ki}$ is the distance between the center of the $k$-th bead and the center of the $i$-th cell. After calculating the probabilities for all cells and side-chain beads, they are normalized so that $\hat{p}_{ki}\equiv p_{ki} / \sum_k p_{ki} $, with $\sum_k \hat{p}_{ki} =1$, where the sum is over the beads occupying cell $i$. 
Note that $p_{ki}$ favors smaller beads that are closer to the cell, thereby preventing their disappearance. If the only bead occupying the $i$-th cell is the $k$-th bead, then $\hat{p}_{ki} = 1$.

Next, we apply the same procedure to map the C$\alpha$ beads. For all amino acids except glycine and proline, the C$\alpha$ beads are treated as alanine residues. This choice is based on the assumption that backbone regions are predominantly hydrophobic, similar to alanine. At this stage, only empty grid cells are considered, as side-chain beads always take precedence over C$\alpha$ beads.

This mapping algorithm can be readily adapted to any implicit solvent model, including those for nanoparticles, membranes, surfaces, and beyond. The key requirement is that the diameter of the building units (such as OPEP beads in our study) should be at least as large as or larger than the vdW diameter of a water molecule, $r_0$. If this condition is not met, the lattice resolution may become too coarse, limiting the algorithm's ability to accurately represent the building units.

To build the coupled CVF-OPEP simulation, empty cells are assigned to contain CVF water. Therefore, the total number of cells is ${\cal N} \equiv N_{\rm w} + N_{\rm R}$, where $N_{\rm w}$ is the number of water molecules and $N_{\rm R}$, the number of cells that are occupied by the proteins. 

The Hamiltonian of the hydrated proteins is
\begin{equation}
    \label{eq:CVF-protein-hamiltonian}
    \mathcal{H} \equiv \mathcal{H}_{\rm bulk} + \mathcal{H}_{\rm hyd} + \mathcal{H}_{\rm R,w} + \mathcal{H}_{\rm R,R},
\end{equation}
where the right-hand terms are the Hamiltonian for bulk CVF water, Eq.~(\ref{ham_bulk}), the water within the hydration layer at the interface with biomolecules, the residues interacting with hydration water, and the residues interacting with other residues, respectively. 

The $\mathcal{H}_{\rm hyd}$ and the corresponding enthalpy have the same form as Eq.~(\ref{ham_bulk}) and Eq.~(\ref{eq:enthalpy}), respectively, but with $J$, $J_\sigma$, and $v_{\rm HB}$ replaced by parameters that account for the effect of the hydrated protein interfaces \cite{Moelbert2003, Petersen2009, Tarasevich2011, Dill2005, Bischofberger2014, Badasyan2015, Fenell2011, Camilloni2016, BiancoPRX2017}. These interface parameters are, respectively, $J^\Theta_l$, $J^\Theta_{\sigma,l}$, and $v^\Theta_{\rm HB, l}$, where $\Theta$ can have three values depending on the properties of the residues near the interacting hydration molecules: PHO for hydrophobic, PHI for hydrophilic, and MIX if one residue is hydrophobic and the other is hydrophilic.

The index $l\in \{1, n_l\}$ indicates the layer in the hydration shell with $n_l=3$, consistent with recent all-atom MD studies \cite{Martelli:2021wp}.
The first layer consists of water molecules occupying the first and second neighboring cells of proteins exposed to the solvent (Fig.~\ref{fig:hyd_shell_def}). The subsequent $i$-th layer is similarly defined as the first and second neighbors of water molecules in the ($i-1$)-th layer. 

For example, a HB between two water molecules within the first hydration layer near two PHO residues will reduce the system's total energy by $J^{\rm PHO}_1$ and increase the total volume by $v^{\rm PHO}_{{\rm HB}, 1}$. Their many-body contribution will have the characteristic energy $J^{\rm PHO}_{\sigma, 1}$. The setting of these parameters is discussed later in this section.

The residue-water interaction is controlled by  
\begin{equation}
    \label{eq:CVF-protein_ham_Rw}
    \mathcal{H}_{\rm R,w} \equiv \sum_k^{N_{\rm R}} \left[ \sum_{i}^{N_{\rm w}} C_{ki} S^{\rm w}_{k} \right],
\end{equation}
where the outer sum runs over all the $N_{\rm R}$ residues in the system, the inner sum over all the $N_{\rm w}$ water molecules, the index $C_{ki}$ is 1 if the residue $k$ and the water $i$ are in nearest neighbors cells, 0 otherwise, and the
parameter $S^{\rm w}_{k}$ is the water-residue interaction energy that depends on the residue $k$ properties and is discussed later in this section.

The residue-residue interaction term, $\mathcal{H}_{\rm R, R}$, generally mirrors the form of $\mathcal{H}_{\rm R,w}$, utilizing standard residue-residue energy scales for neighboring residues, as described in Ref.\cite{BiancoPRX2017}. This interaction is especially significant when residues are free to move and their configurations evolve along the MC trajectories. Within the implicit-solvent framework of the OPEP force field, both inter- and intra-protein energy interactions are considered, affecting the initial protein configurations that we hydrate at the start of our water-explicit simulations. Since our trajectories are generated with fixed protein configurations, we set $\mathcal{H}_{\rm R, R} \equiv 0$ to avoid double-counting the residue-residue contribution to the total enthalpy. 


Setting the model's parameters at the interface is complex and requires a multiscale approach, which is beyond the scope of this work. Our main goal here is to qualitatively demonstrate the feasibility and usefulness of coupling CVF water with implicit-solvent protein configurations. 
For the purposes of this work, we adopt the classification of OPEP amino acids into non-polar, polar, and charged \cite{Samanta2021}. We consider polar and charged beads as hydrophilic, while non-polar are hydrophobic \cite{Kuriyan:2013aa}. 
We assume that water interacts with hydrophobic residues only by excluded volume. Hence, $S^{\rm w}_{k}=0$ if the residue is hydrophobic. Also, we simplify the water interaction with the hydrophilic residues by considering that the corresponding hydration-energy gain is approximately independent of the amino acid. 
Since fully-exposed larger residues have more hydration water, we set $S^{\rm w}_{k}/(4\epsilon_0)=-0.1$ if the radius of the bead is larger than the water vdW diameter, i.e., $r_k\geq r_0$, and $S^{\rm w}_{k}/(4\epsilon_0)=-0.5$ otherwise (Table~\ref{table:amino_acids}). 

Furthermore, for simplicity, for the water within the hydration shell, we adopt the same parameters used in similar qualitative works limited to proteins in water monolayers \cite{BiancoPRX2017, CoronasBook2022} (Table~\ref{table:CVFExtended_parameters_BiancoPRX}) as discussed in Appendix~\ref{si:cvf_parameters}.

\subsection{Simulation method}
\label{simulation_method}

To investigate the sequestration of folded SOD1 in highly concentrated solutions of BSA and FUS, we solvate the configurations sampled via the OPEPv4 model and LBMD \cite{Samanta2021} with CVF water, as detailed in Sec.~\ref{couple_cvf_opep}. Since water DoF relax faster than the characteristic timescale of amino acid motions, we fix each OPEP configuration and generate a MC trajectory to equilibrate water. 

Note that there is no re-weighting of OPEP configurations, so the likelihood of observing specific states remains the same. Additionally, the dynamics between these states are derived from the original implicit-solvent LBMD simulations and are not affected by the presence of explicit solvent. Instead, each hydrated OPEP configuration acts as the starting point for a new trajectory, during which water reaches equilibrium. By analyzing these trajectories, we can examine water-related observables, which then enable us to project the free energy landscape onto water-dependent coordinates. 

\begin{figure*}[]
    \centering
    \includegraphics[scale=0.28]{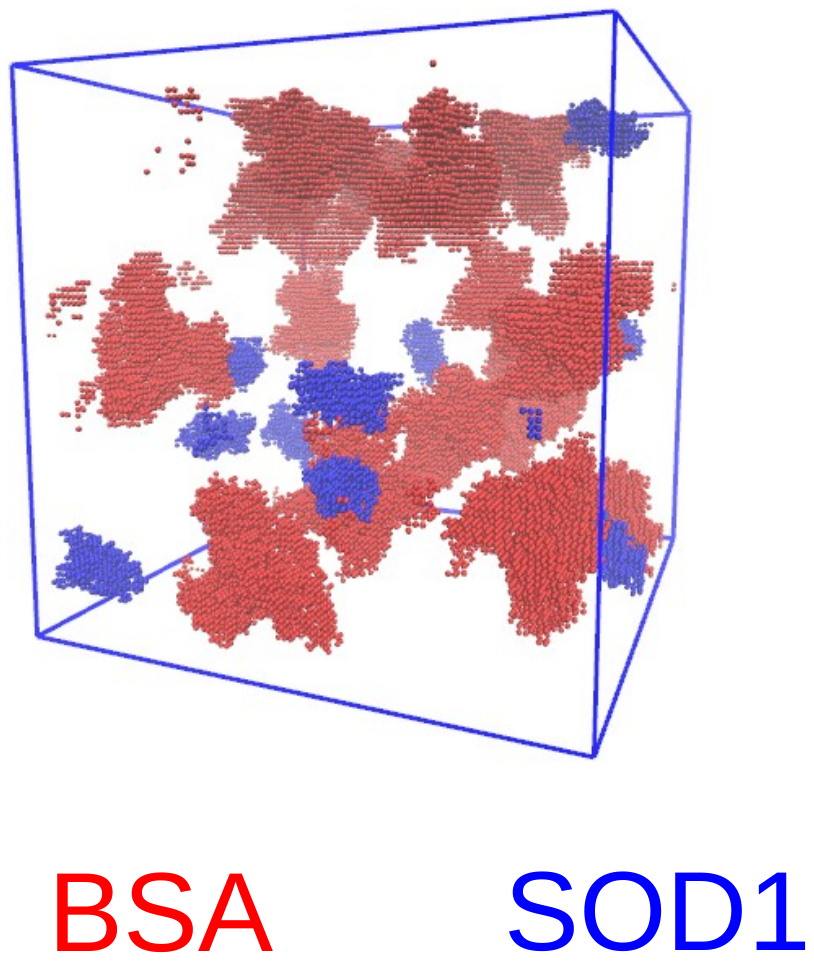}
    \includegraphics[scale=0.28]{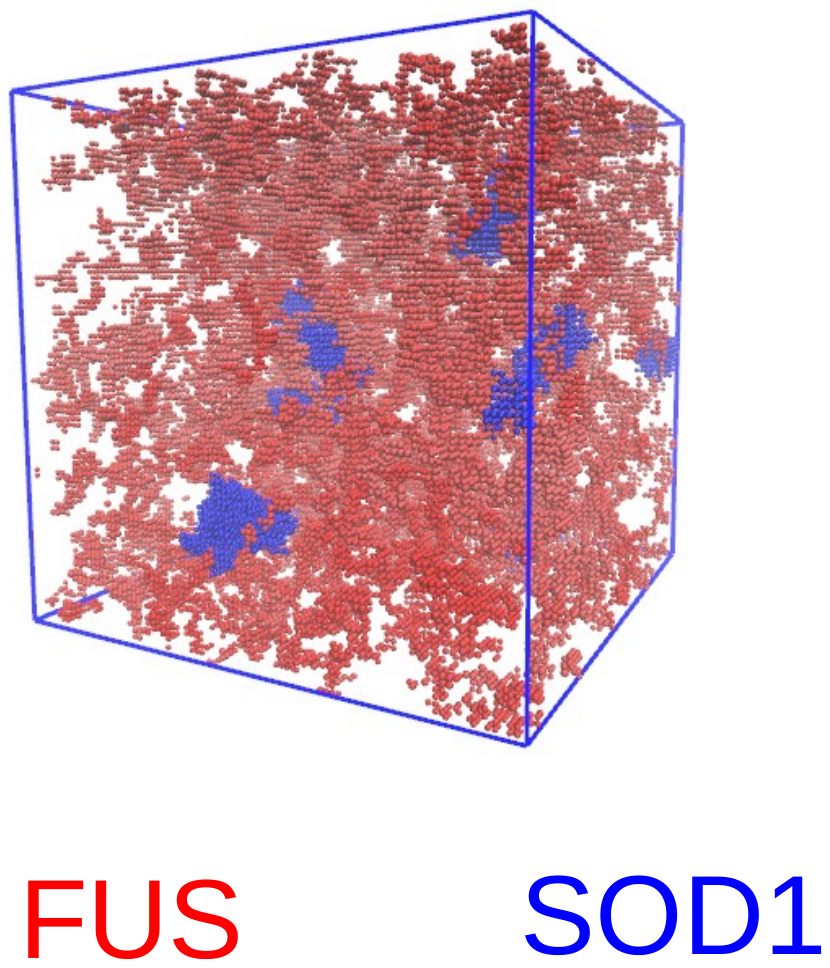}
    \caption{OPEP-dis configurations for SOD1 sequestration in BSA (left) and FUS (right) crowded environments with water not shown. Blue beads represent cells occupied by SOD1, while red beads indicate cells occupied by the crowder. Water molecules (not shown) fill all the available volume left by the proteins. Blue lines mark the boundaries of the simulation box (255~\AA), where pbc apply. }
    \label{fig:sod1_simulationBox}
\end{figure*}

We set the temperature and pressure to $T=300{\rm ~K}$, $P=1{\rm ~atm}$ to match those of the original work \cite{Samanta2021}. We divide a cubic volume with a 255~\AA~ lateral side into ${\cal N} = 681,472$ cells (88 cells per side) and apply periodic boundary conditions (pbc) in all directions. 
The LBMD-OPEP simulations for SOD1 with BSA and FUS last about 1 $\mu$s, with configurations saved every 0.2 and 5 ns, respectively, and have comparable concentrations: 10 SOD1 in 15 BSA proteins (100 g/L) and 10 SOD1 in 70 FUS chains (150 g/L) \cite{Samanta2021}.
We convert those roughly 5,000 and 200 protein states into OPEP-dis configurations, solvate them with $N_{\rm w} \sim 630,000$ water molecules, and equilibrate the solvent using Metropolis MC in the $NPT$ ensemble, parallelized on GPUs \cite{coronas2025algorithms}. For each fixed protein configuration, we sample over 10,000 independent water configurations (Fig.~\ref{fig:sod1_simulationBox}).

\subsection{Free energy calculations}
\label{method:free_energy}

If $\vec{q}$ are generalized coordinates that define the macroscopic state (macrostate) of the system, and ${\cal P}(\vec{q})$ is the probability density of finding the system in the macrostate $\vec{q}$, the Gibbs free energy of the macrostate is
\begin{equation}
\Delta G(\vec{q}) \equiv -k_BT \log\left({\cal P}(\vec{q})\right),
\end{equation}
where $k_B$ is the Boltzmann constant.
Note that a macrostate comprises many protein configurations and can represent a specific associative state between two or more proteins. A macrostate is a thermodynamic concept defined by macroscopic quantities $\vec{q}$, while a state is a microscopic entity defined by molecular coordinates. 

To determine whether including solvent DoF in the free energy landscape provides valuable insights into the properties of the macrostate, we analyze two sets of generalized coordinates: one derived from implicit-solvent OPEP simulations and another that encompasses both protein and solvent coordinates. 
When the macrostate coordinates $\vec{q}$ are independent of the water DoF, the free energy $\Delta G(\vec{q})$ does not depend on the CVF model's parameter choices. However, if the macrostate coordinates $\vec{q}$ are water-dependent, since the water DoF are equilibrated based on the parameters discussed in Sec.~\ref{couple_cvf_opep}, then the projection $\Delta G(\vec{q})$ depends on the CVF model's parameters. Nevertheless, as discussed in \href{./supplementary.pdf}{Supplementary information: Selection of water parameters within the hydration shell}, the results are consistent despite variations in these parameters.

\section{Results}
\label{results}

\subsection{Hydration of SOD1 in BSA and FUS crowders}
\label{results:hydration}

\begin{figure*}[]
    \centering
    \includegraphics[scale=0.45]{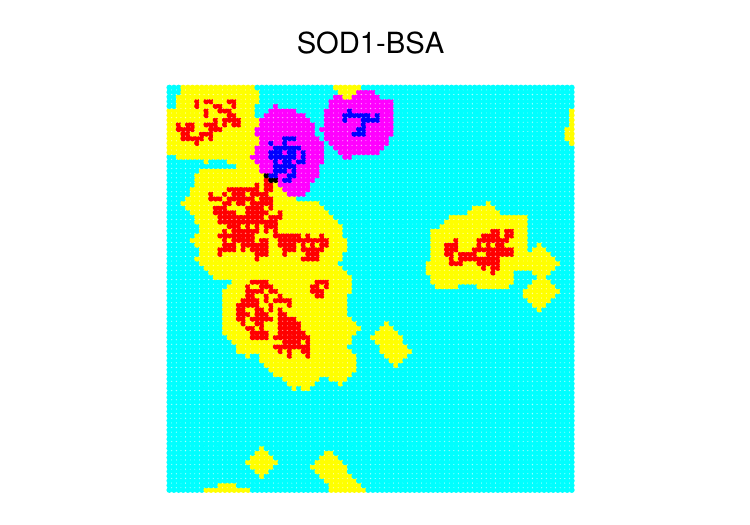}
    \includegraphics[scale=0.45]{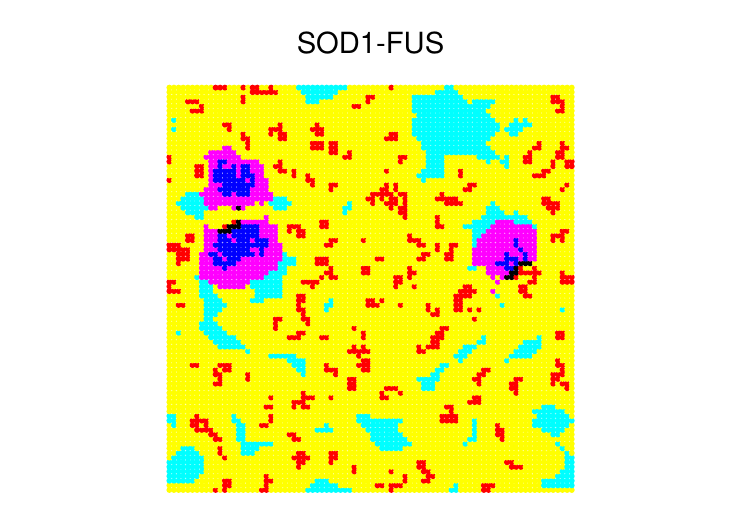}
    \caption{OPEP-dis configurations for SOD1 sequestration in BSA (left) and FUS (right) crowded environments with water explicitly shown. 
For clarity, we represent a 2D section of the 3D system. 
    Each volume cell is represented by a colored point: Cyan for bulk water; Blue for SOD1; Red for CWD; Magenta for water hydrating SOD1; Yellow for water hydrating CWD; Black for water hydrating both (mixed category).}
    \label{fig:sod1_hydrationLayer}
\end{figure*}

The $N_{\rm w}$ water molecules are within the hydration shell or in bulk, i.e., $N_{\rm w} \equiv N_{\rm hyd} + N_{\rm bulk}$, and the first category can hydrate SOD1, (BSA or FUS) CWD, or both (mixed category). In the mixed case, water molecules are at the same distance from both SOD1 and CWD. Hence, $N_{\rm hyd} \equiv N_{\rm hyd, SOD1}+ N_{\rm hyd, CWD}+ N_{\rm hyd, mix}$ (Fig.~\ref{fig:sod1_hydrationLayer}).

The analysis of water distribution through eye inspection reveals significant variation between hydration and bulk depending on the crowder used. In the case of SOD1 within BSA, the predominant water component is bulk water. Conversely, in FUS, a majority of the water hydrates FUS chains. This variation stems from the distinct conformational characteristics of the crowders: BSA is a globular protein, whereas FUS is intrinsically disordered and highly flexible. Consequently, FUS residues are more evenly distributed throughout the system, enhancing their accessibility to solvent molecules.

\subsection{Hydration enthalpy calculations}
\label{results:enthalpy}

We calculate the total enthalpy of the system as
\begin{equation}
H \equiv H_{\rm bulk} + H_{\rm hyd},
\end{equation}
where $H_{\rm bulk}$ is the enthalpy of bulk water, Eq.~(\ref{eq:enthalpy}), and $H_{\rm hyd}$ is the enthalpy of the hydration shell, including HBs at the hydration layer and water-protein interaction energy:
\begin{equation}
H_{\rm hyd} \equiv \mathcal{H}_{\rm R,w} + \sum_{\Theta,l} \left[ -J^\Theta_{l} N^\Theta_{{\rm HB}, l} - J^\Theta_{\sigma,l} N^\Theta_{\sigma, l} - P v^\Theta_{{\rm HB},l} N^\Theta_{{\rm HB},l} \right].
\end{equation}
This expression accounts for the setting $\mathcal{H}_{\rm R,R} =0$ discussed 
in Sec.~\ref{cvf}. 

Despite differences in water partitioning between bulk and hydration layers, our overall estimate of the total enthalpy $H$ remains similar for both cases, with the SOD1 sequestration into FUS being approximately 0.6 kJ/mol more advantageous. However, when considering only the hydration shell component $H_{\rm hyd}$, the difference widens to about 2 kJ/mol. This discrepancy in $H_{\rm hyd}$ can be attributed to the fact that the FUS system contains more water molecules in its hydration shell, and its proteins are more exposed to solvent. Overall, most of the enthalpy difference in the hydration shell is offset by a larger $H_{\rm bulk}$ component in the BSA system, balancing the energetics between the two (Table~\ref{table:sod1_enthalpy_resutls}).

\subsection{Free energy landscape}
\label{results:free_energy}

\subsubsection{Implicit-solvent analysis}

To define a set of relevant generalized coordinates $\vec{q}$ that adequately describe the macrostate of the system, for each SOD1 $i$ and OPEP-dis configuration at time $t$, we define the adsorption profile as $c_i(d,t) \equiv N_i(d,t)/N_i(t)$, where $N_i(d,t)$ is the number of cells of the $i$-th SOD1 exposed to the solvent that are at distance $d$ from the CWD, and $N_i(t)$ is the total number of cells of the $i$-th SOD1 exposed to the solvent at time $t$. The physical meaning of $c_i(d,t)$ is the fraction of the $i$-th SOD1 surface that, at time $t$, is at the distance $d$ from the CWD. 

We find that the average over $t$ and all the SOD1 proteins, $c(d)\equiv \langle c_i(d,t) \rangle$, reflects the different conformational properties of the CWDs (Fig.~\ref{fig:sod1_adsorption_profile}). In particular, $c(d)$ of SOD1 in FUS exhibits a single peak at short distances $d$, with a sharp maximum at the first hydration shell, while SOD1 in BSA has a broad adsorption profile with a maximum at the second hydration shell. We then define the adsorption factor as the integral
\begin{equation}
\label{eq:sod1_adsorption_factor}
   {\cal C}_i(t) \equiv \int d \cdot c_i(d,t)  \cdot {\rm d}d.
\end{equation}
Hence, ${\cal C}_i(t)$  is the average distance between the surface of the $i$-th SOD1 and the surface of the CWDs at the time $t$. We choose ${\cal C}_i$ as the first component of the generalized coordinates $\vec{q}$ that we use to describe the macostate of the system.

\begin{figure*}[]
    \centering
    \includegraphics[scale=0.4]{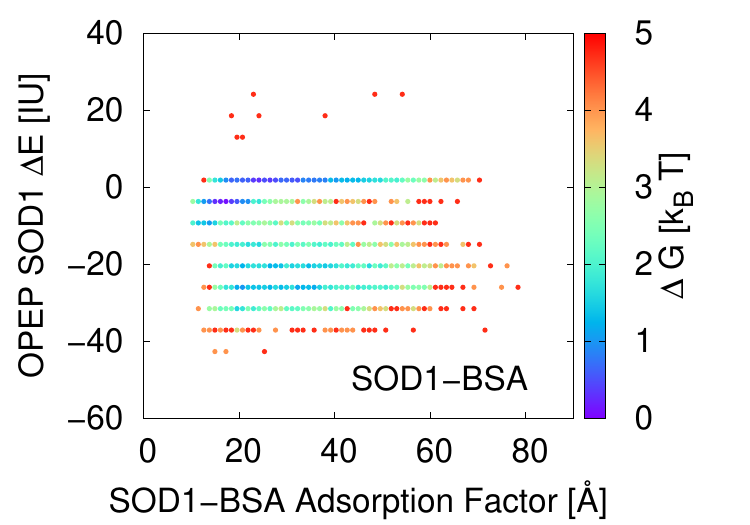} 
    \includegraphics[scale=0.4]{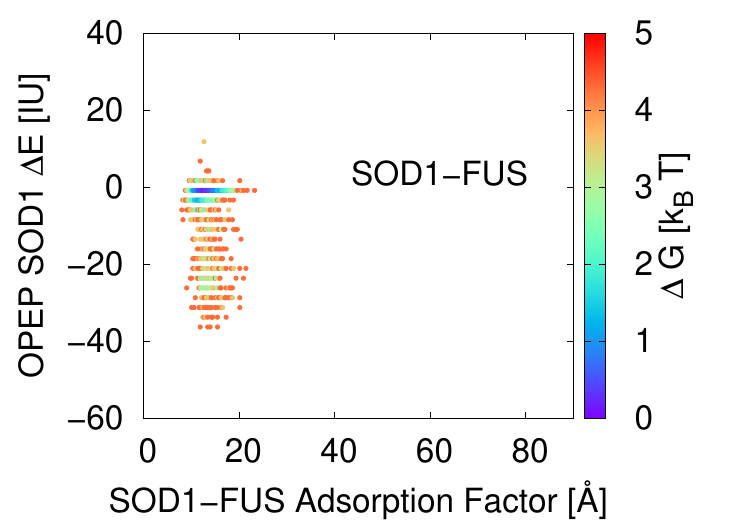} \\
    \includegraphics[scale=0.4]{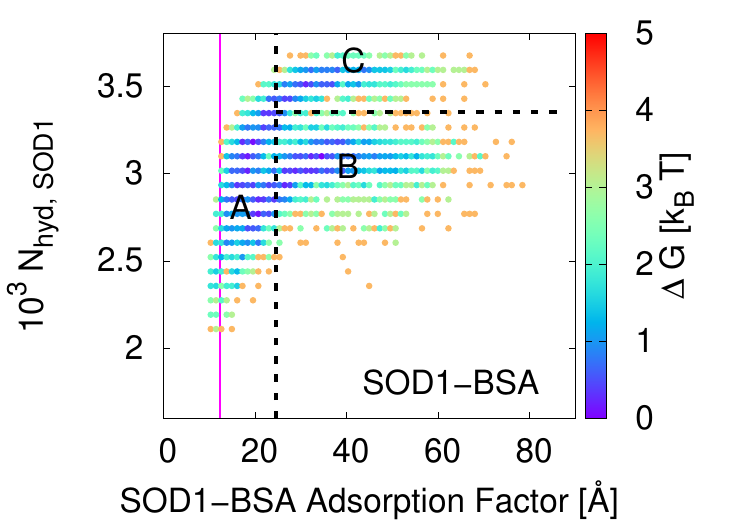}
    \includegraphics[scale=0.4]{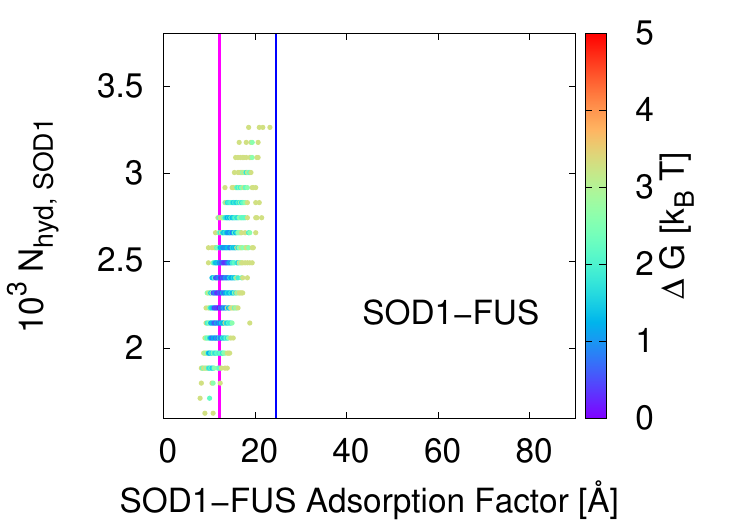} 
    \caption{Free energy $\Delta G$ landscape of SOD1 into BSA (left panels) and FUS (right panels) solutions. Top panels: $\Delta G\left(\Delta E_i, {\cal C}_i\right)$, calculated from implicit solvent simulations. Bottom panels: $\Delta G \left(\Delta N_{{\rm hyd, SOD1},i}, {\cal C}_i \right)$, calculated from explicit solvent simulations. Magenta and blue lines indicate the distances of one and two hydration shells, $d_{\rm Hyd.~shell}$ and $2d_{\rm Hyd.~shell}$, respectively. Lower-left panel (SOD1-BSA in explicit solvent): Black dashed lines separate the landscape into three regions corresponding to A, B, and C states, as described in the text. The vertical dashed line, separating A from B and C, is at ${\cal C}=2d_{\rm Hyd.~shell}$. The value of $\Delta G$ is color-coded, as shown in the bars on the right of each panel.}  
    \label{fig:sod1_freeEnergy}
\end{figure*}

For the implicit solvent case, we define the second component of $\vec{q}$ 
as the interaction energy of the $i$-th SOD1 with all other proteins in the system, $\Delta E_i$, calculated using the OPEP force field. We then plot the free energy $\Delta G = -k_B T \log\left( {\cal P} ({\cal C}_i, \Delta E_i) \right)$ (Fig.~\ref{fig:sod1_freeEnergy}, top panels).

The free energy landscape projected onto $\vec{q}=({\cal C}_i, \Delta E_i)$ for SOD1 in BSA (Fig.~\ref{fig:sod1_freeEnergy}, top-left panel) reveals a prominent basin with a minimum around $\Delta E \simeq 0$. Within this minimum, most states have ${\cal C}_i > 12$~\AA, indicating that SOD1 generally tends to be positioned at a distance greater than the non-bonded cutoff of the OPEP force field, which is $10$~\AA. However, some states within the same minimum exhibit $\Delta E < 0$ and ${\cal C}_i \lesssim 10$~\AA, suggesting states where SOD1 is in close proximity to BSA, resulting in non-zero interaction energy. Additionally, many states with $\Delta E < 0$ and ${\cal C}_i > 12$~\AA~  are accessible through thermal fluctuations at less than $1 k_BT$ of free-energy cost, representing likely SOD1-SOD1 contacts away from BSA. Overall, the free energy landscape appears relatively flat and does not distinctly differentiate between various states of SOD1 adsorption onto BSA.

For SOD1 in FUS (Fig.~\ref{fig:sod1_freeEnergy}, top-right panel), the free energy landscape projected onto $\vec{q}=({\cal C}_i, \Delta E_i)$ shows a minimum with $\Delta E \lesssim 0$, separated by a free-energy barrier of about $5k_BT$ from a less populated metastable basin at $\Delta E < -20$ and $\Delta G \simeq 3k_BT$, with both states at ${\cal C}_i \lesssim 12$~\AA. Since FUS proteins are more uniformly distributed in the system, SOD1s are always close. However, the weakly interacting state, with $\Delta E \lesssim 0$, appears more stable than the strongly interacting state, with $\Delta E < -20$.

Therefore, the implicit-solvent analysis suggests minimal interaction between SOD1 and BSA, with a wide range of distances observed between the proteins and the CWDs. It also shows only weak interactions of SOD1 with FUS proteins that are close and evenly distributed. Hence, the primary difference between the two cases is the volume accessible to the SOD1.

\subsubsection{Explicit-solvent analysis}
\label{explicit}

Next, we demonstrate that the picture obtained from including water explicitly in the free-energy analysis offers more detailed insights, making it essential for accurately representing the system's macrostate. 
To this end, we consider a second set of generalized coordinates $\vec{q}=({\cal C}_i, N_{{\rm hyd, SOD1}, i})$, where $N_{{\rm hyd, SOD1}, i}$ is the number of water molecules hydrating the $i$-th SOD1. Therefore, we aim to project the free energy onto a plane where the first coordinate represents information about the spatial distribution of the solutes, and the second relates to the solvent and the hydration state of the SOD1. 
Using this projection, we can study a landscape that allows us to gain deeper insights into the free energy of SOD1 within the two CWDs. This enhances the previous implicit-solvent study and provides a clearer understanding of the physical mechanisms responsible for SOD1's sequestration (Fig.~\ref{fig:sod1_freeEnergy} bottom panels).

For SOD1s in the BSA solution (Fig.~\ref{fig:sod1_freeEnergy} bottom-left panel), the free energy landscape displays three characteristic macrostates connected within a single basin. The three macrostates correspond to three different associative states of SOD1 with BSA. We designate these states as 'A', 'B', and 'C'.  

\begin{figure*}[]
    \centering
    \includegraphics[trim=3cm 5cm 0cm 5cm,clip,width=0.31\textwidth]{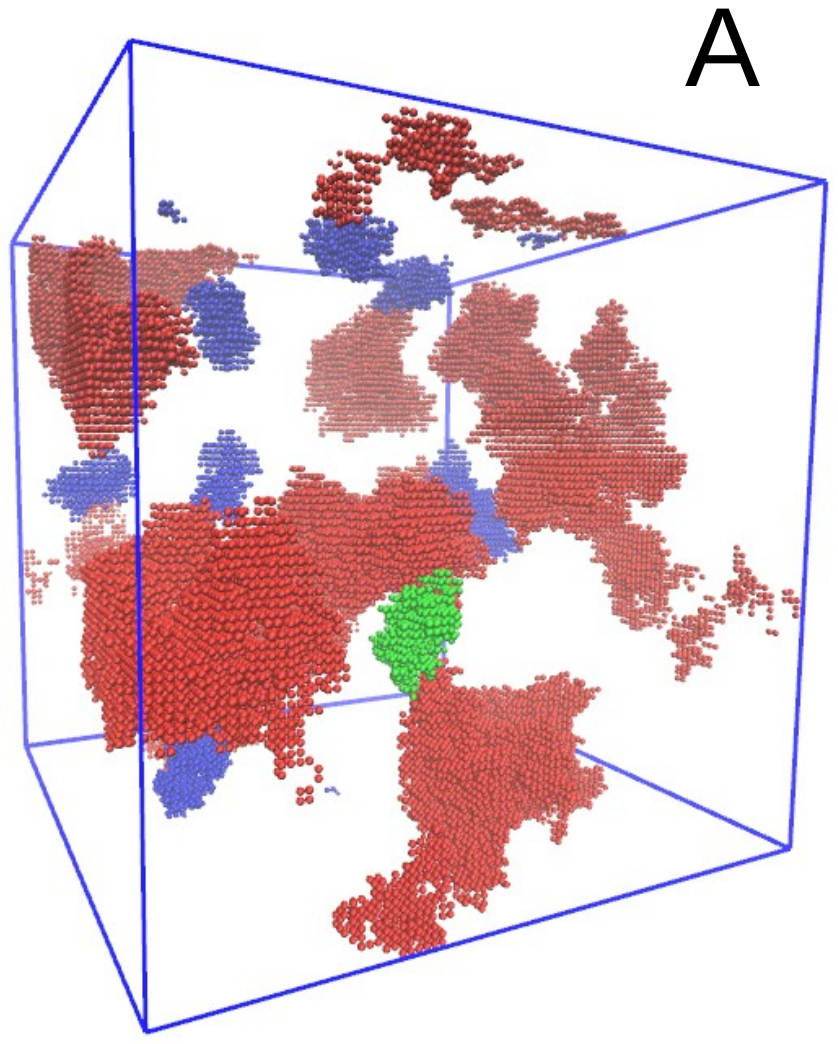}
    \includegraphics[trim=3cm 5cm 0cm 5cm,clip,width=0.31\textwidth]{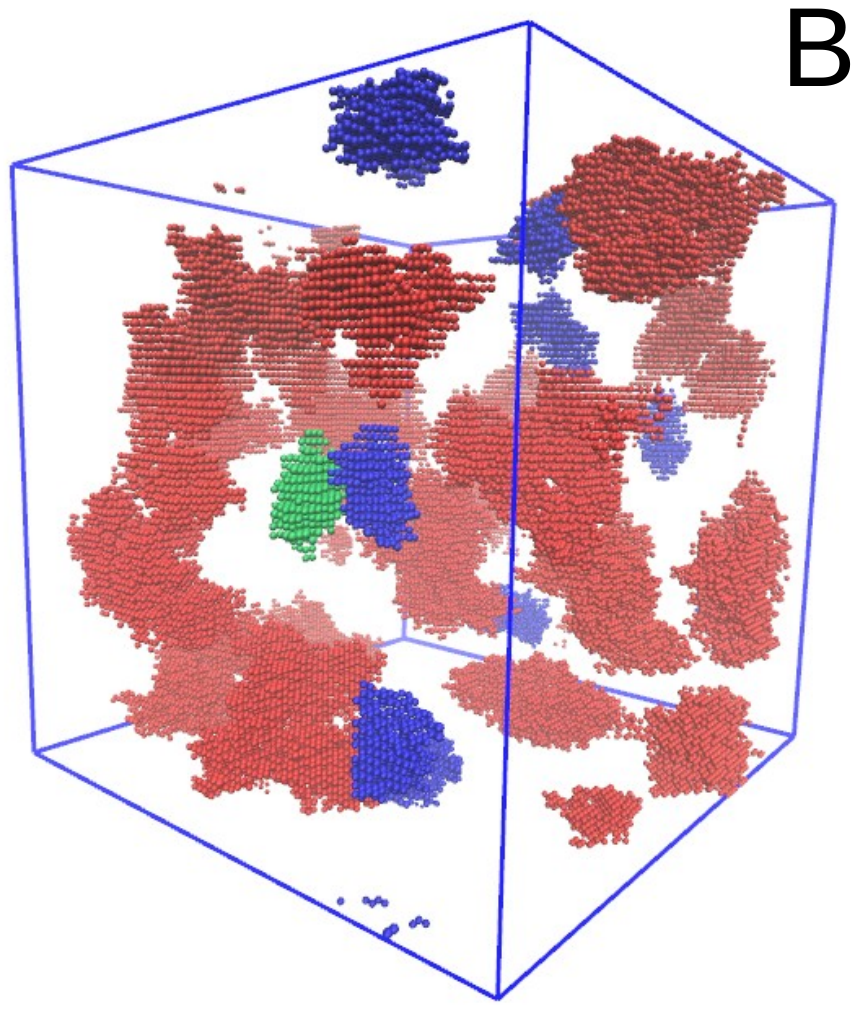}
    \includegraphics[trim=1.8cm 5cm 0cm 5cm,clip,width=0.31\textwidth]{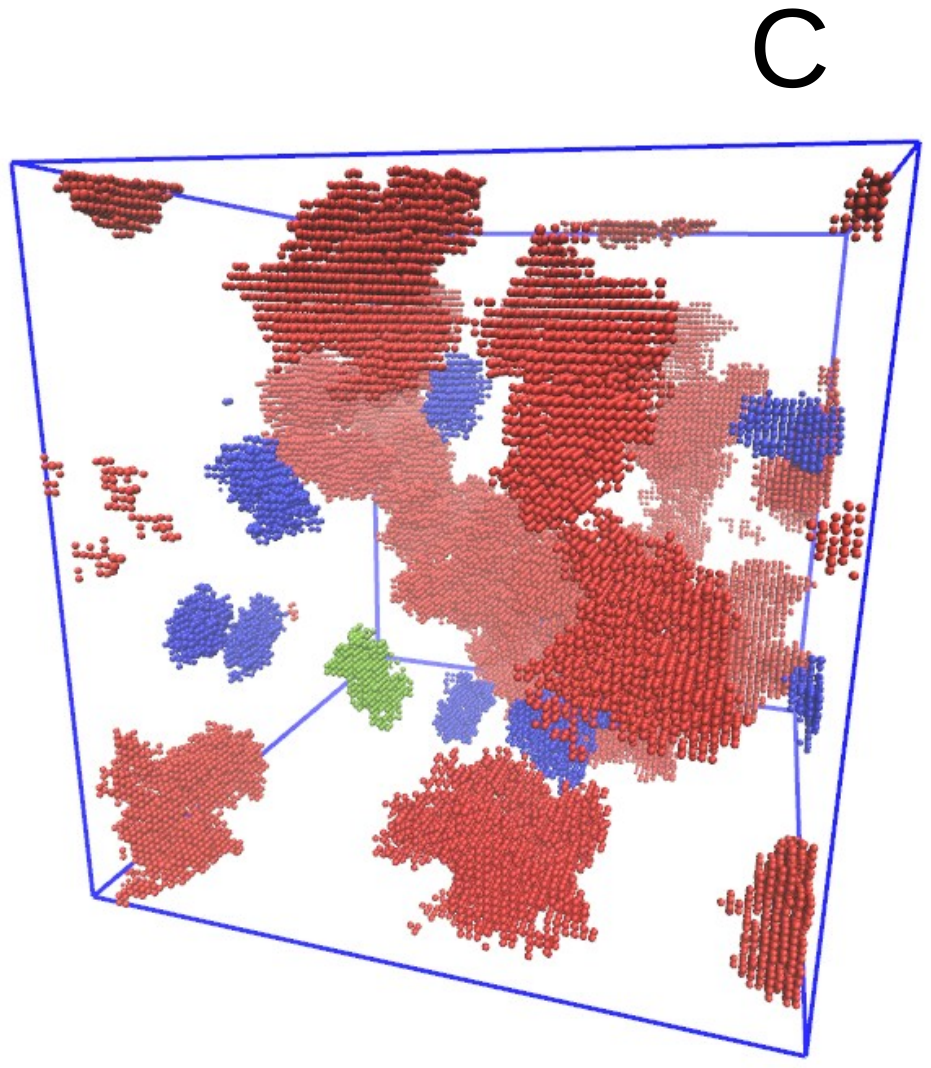}
    \caption{Typical configurations of associative states A, B, C for SOD1 in BSA crowders. The states are defined for individual SOD1 (in green). The other SOD1 proteins are in blue, and the BSA proteins are in red. In each configuration, water molecules (not shown) fill all the available volume left by the proteins. Blue lines mark the boundaries of the simulation box (255 \AA) with pbc.
    Left: State A, with SOD1 in contact with BSA. 
    Center: State B, with SOD1-SOD1 contact.
    Right: State C, with SOD1 away from other proteins.}  
    \label{fig:sod1_BSA_configurations}
\end{figure*}

The state A comprises configurations with adsorption factors less than two hydration shells, ${\cal C}_i < 2\cdot d_{\rm Hyd.~shell} \sim 25$~\AA~ (six water layers), where $d_{\rm hyd.~shell} \equiv \sqrt{(n_l\cdot r_0)^2+(n_l\cdot r_0)^2}=4.24\cdot r_0$, because each water layer includes up to second neighbors. At these short ${\cal C}_i$ values, the hydration shells of SOD1 and BSA overlap. Therefore, in state A, SOD1 and BSA have either direct or water-mediated interactions (Fig.\ref{fig:sod1_BSA_configurations}.A).

For larger ${\cal C}_i$, we identify two states (B and C) separated by a free energy barrier of $\simeq 3 k_BT$. The B and C states differ in the number of water molecules in the hydration shell. In state C, $N_{{\rm hyd, SOD1}, i} > 3,350$ molecules, indicating that the SOD1 is surrounded by a complete hydration shell and is far from any other protein (Fig.\ref{fig:sod1_BSA_configurations}.C).

In contrast, for state B, $N_{{\rm hyd, SOD1}, i}$ is approximately 3,000. This indicates that the hydration shell of SOD1 overlaps with that of another protein. Given that the average distance from BSA is large, with ${\cal C}_i > 2d_{\rm Hyd.~shell}$, the state B consists of configurations where at least two SOD1 molecules are interacting with overlapping hydration shells (Fig.\ref{fig:sod1_BSA_configurations}.B).

Note that these three states are difficult to distinguish in the implicit-solvent free energy projection (Fig.~\ref{fig:sod1_freeEnergy} top-left panel). Although the multidimensional free energy is, by design, the same in both the top-left and bottom-left panels of Fig.~\ref{fig:sod1_freeEnergy}, the two landscapes are complementary when changing one generalized coordinate. In the implicit-solvent projection, hydration shells cannot be defined, and state A encompasses all possible $\Delta E$ values within an arbitrary radius of about 25 \AA.
In contrast, states B and C represent all configurations beyond roughly 25 \AA, with $\Delta E$ ranging from less than zero for state B to approximately zero for state C. However, there is little to no free energy barrier between these configurations. Therefore, the implicit-solvent landscape is so flat that distinguishing states without including the solvent-dependent generalized coordinate is challenging, if not impossible.

In the hydration-projection of the free energy of SOD1 within FUS (Fig.~\ref{fig:sod1_freeEnergy} bottom-right panel), we identify a single basin with a minimum with SOD1 mainly within the first hydration shell of FUS and an average hydration level between roughly 2,000 and 3,000 water molecules. 
Therefore, the suggested distinction between weakly- and strongly-interacting SOD1-FUS configurations, based on the implicit-solvent analysis (Fig.~\ref{fig:sod1_freeEnergy} top-right panel), is better understood as a difference between more-hydrated and less-hydrated SOD1 interacting with FUS, respectively, all within the same free-energy basin.

\subsection{Kinetic behavior of SOD1 in BSA crowders}

\begin{figure}[]
    \centering
    \includegraphics[scale=0.22]{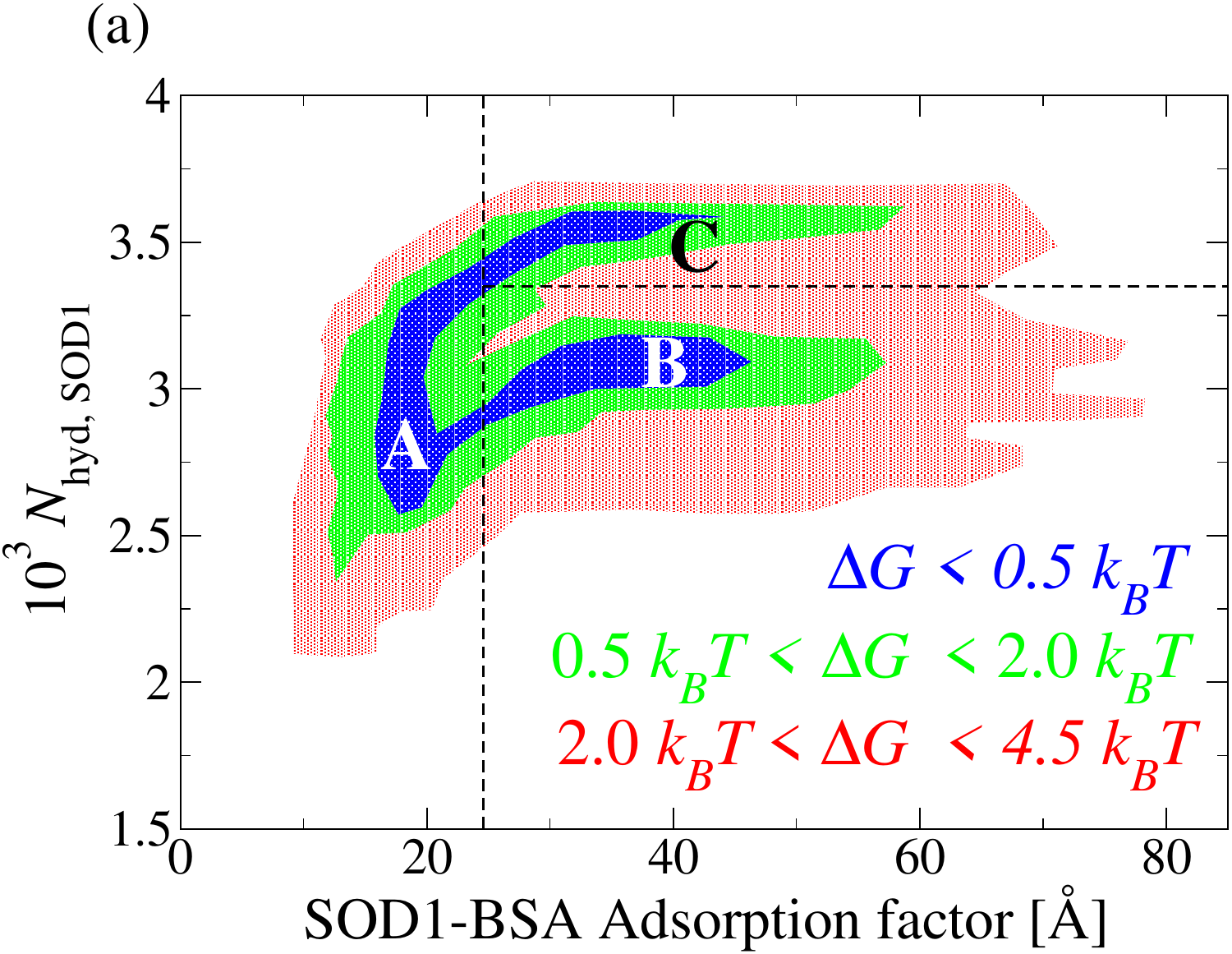}
    \includegraphics[scale=0.46]{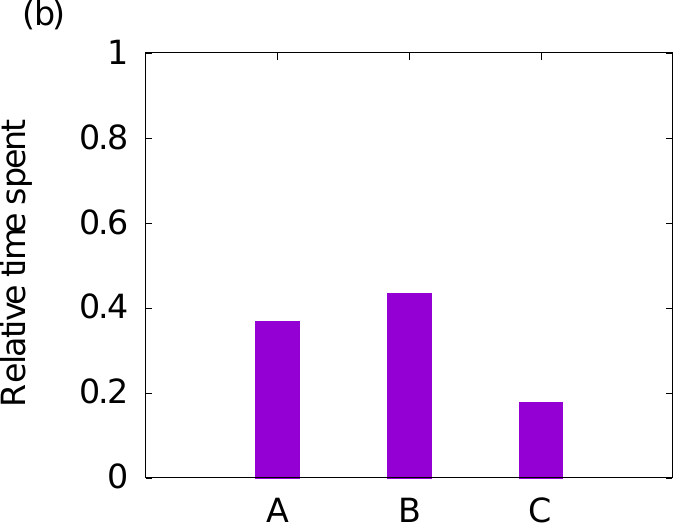}
    \includegraphics[scale=0.46]{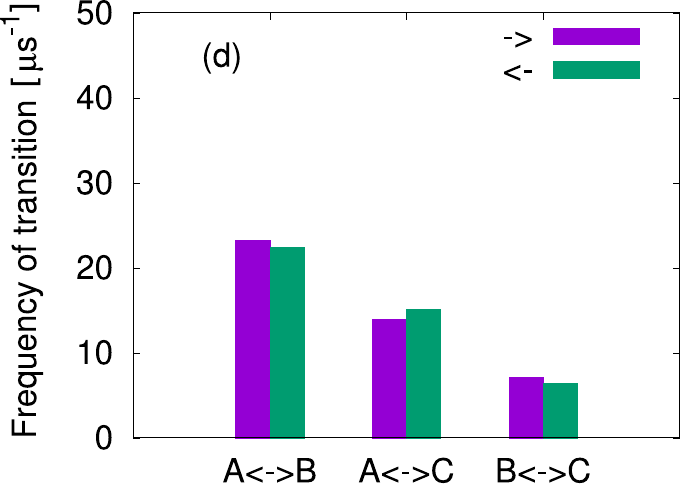}
    \includegraphics[scale=0.22]{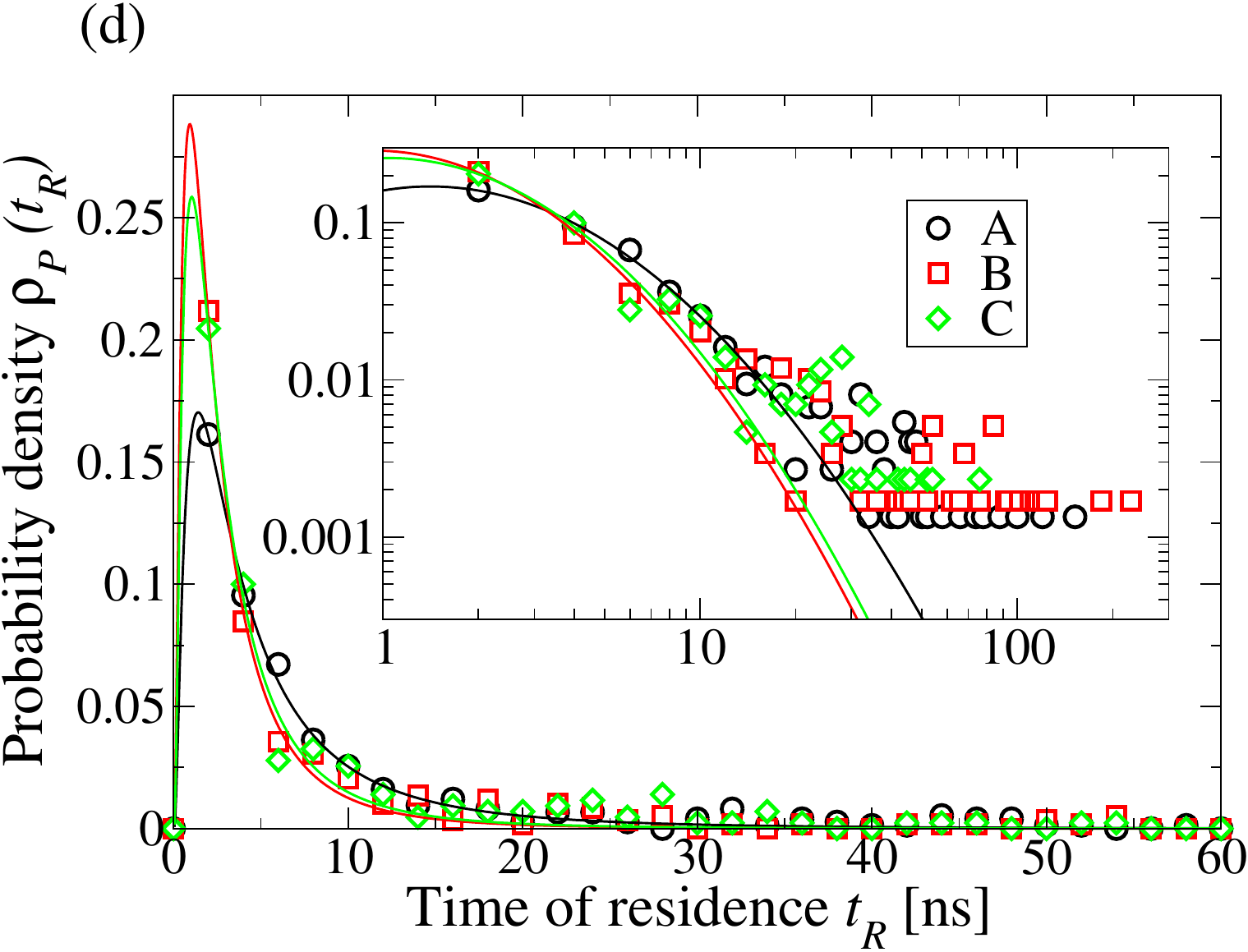}
    \caption{Kinetcs analysis of SOD1 in BSA crowders. 
    (a): Schematic hydration free energy landscape for SOD1 in the BSA solution. The blue region corresponds to  $\Delta G /k_BT\lesssim 0.5$, green to  $0.5 \lesssim \Delta G/k_BT \lesssim 2$, and red to $2 \lesssim \Delta G/k_BT \lesssim 4.5$. A, B, and C labels denote the associative states described in Fig.~\ref{fig:sod1_BSA_configurations}. 
    (b): Time the system spends in each of the three states relative to the total simulation time. 
    (c): Frequency of transitions among the three states, measured as the number of transitions per unit of time.  
    (d): Probability density $\rho_P(t_R)$ of SOD1 residence time $t_R$ in states A (black circles), B (red squares), or C (green diamonds), with log-normal distribution fits (lines with matching colors; with fitting parameters in Table~\ref{table:lognorm_fit}).  
Results are averages of data in Figs. (\ref{fig:transition_residence_0_1}--\ref{fig:transition_residence_8_9}).}
    \label{fig:sod1_residence_transitions}
\end{figure}

When water is explicitly considered in the analysis, the free energy landscape of SOD1 within BSA crowders can be described by three associative states: A, B, and C (Fig.~\ref{fig:sod1_residence_transitions}.a). These states are interconnected, with state A positioned between B and C, acting as an intermediary. This framework enables the examination of the population kinetics of protein configurations transitioning among these three states. 

First, we estimated the proportion of time that SOD1 spends in each state A, B, and C (Fig.~\ref{fig:sod1_residence_transitions}.b). Our findings indicate that SOD1 preferentially interacts with other SOD1 proteins (state B) and to a lesser extent with BSA (state A), compared to being fully hydrated (state C). This pattern aligns with the understanding that protein-protein interactions reduce enthalpy, thereby favoring states A and B, while also displacing hydration water in the bulk, which increases the system's overall entropy. Consequently, the free energy of the system is primarily influenced by protein-protein attraction and the entropy of the bulk water.

Second, we estimate the frequency of transitions between the different states $\nu_{X\to Y}$ as the number of transitions per unit time, where "X" and "Y" represent states A, B, and C (Fig.~\ref{fig:sod1_residence_transitions}.c). Our results indicate that, within statistical error, the transition rates are approximately symmetric ($\nu_{X\to Y} \sim \nu_{Y\to X}$). The highest transition frequency occurs between states A and B, followed by A and C, which aligns with the absence of free energy barriers for these transitions. Conversely, transitions between B and C occur less frequently due to the presence of a free energy barrier, estimated to be $\lesssim 4k_BT$. This barrier can be surmounted thanks to thermal energy, facilitating these transitions.

Third, we estimate the histogram of uninterrupted residence times, $t_R$, representing the duration during which a SOD1 molecule remains in the same associative state (Fig.~\ref{fig:sod1_residence_transitions}.d). Our findings reveal that $t_R$ spans over three orders of magnitude, from a few nanoseconds up to 200 ns, indicating the absence of a characteristic $t_R$. Considering that negative values of $t_R$ are physically meaningless, we opt to fit the data using a log-normal distribution. The probability density function of the log-normal distribution with parameters $\mu$ and $\sigma$ is

\begin{equation}
\label{eq:log_norm}
   f(t_R) = \frac{1}{t_R\sigma\sqrt{2\pi}} \exp\left(-\frac{(\ln(t_R)-\mu )^2}{2\sigma^2}\right).
\end{equation}

We analyze data within the time range $t_R \leq 20{\rm ~ns}$, as histograms from simulations are insufficiently sampled at larger times due to the total simulation duration being $1 ~\mu{\rm s}$. For each associative state, Table~\ref{table:lognorm_fit} lists the fitted parameters of the log-normal distribution, along with the mean and variance of $t_R$, and the decay exponent $\rho_P(t_R) \propto t_R^\alpha$.

Results presented here for the kinetics are averages of the data for each of the ten SOD1 proteins in the system (Figs. \ref{fig:transition_residence_0_1}--\ref{fig:transition_residence_8_9}).
A detailed analysis shows that individual SOD1s describe different trajectories. For example, 1st and 2nd SOD1s spend approximately 60~\% of the time in the state B, while 8th SOD1 spends less than 10~\%. 
A similar distribution of results is shown also for the time evolution of ${\cal C}_i(t)$, $N_{{\rm hyd, SOD1},i}(t)$, and the visited states A, B, and C (Fig. \ref{fig:sod1_XY_timeseries}).

\begin{table}[t!]
\centering
\begin{tabular}{| c | c | c | c | c | c |}
\hline
  &  $\mu$  & $\sigma$ & $\langle t_R / {\rm ns} \rangle $ & Var$(t_R / {\rm ns})$ & $\alpha$ \\
   \hline
 A &  $1.35\pm0.03$  & $1.00\pm 0.03$ & $6.4 \pm 0.4$  & $16 \pm 2$ & $-1.6\pm0.2$ \\
 \hline
 B &  $0.83 \pm 0.09$ & $0.95 \pm 0.03$ & $3.6 \pm 0.2$   &  $4 \pm 2$ & $-1.8\pm0.3$  \\
 \hline
 C & $0.94 \pm 0.11$ & $0.95 \pm 0.05 $ &  $4.0 \pm 0.6$ & $5 \pm 2$ & $-1.6\pm0.2$  \\
\hline
\end{tabular}
\caption{Parameters of log-normal distribution $(\mu,\sigma)$ for the fit $f(t_R)$ in Eq.~(\ref{eq:log_norm}) of the estimated probability density function $\rho_P(t_R)$ for the three associative states A, B, and C (Fig.\ref{fig:sod1_residence_transitions}.d), including the expected value $\langle t_R\rangle = \exp\left( \mu + \sigma^2/2 \right)$ and the variance ${\rm Var}(t_R) = (\exp\left( \sigma^2\right) -1 )\cdot(\exp\left( 2\mu+\sigma^2\right))$ of $t_R$ in ns, and the exponent $\alpha$ of the power law decay $\rho_P(t_R) \propto t_R^\alpha$. We fit the data in the range $2 {\rm ns}\leq t_R \leq 20 {\rm ns}$.}
\label{table:lognorm_fit}
\end{table}

\section{Discussion}
\label{OPEP:discussion_sod1}

In Ref. \cite{Samanta2021}, Samanta et al. conducted {\it in vitro} experiments showing that the protein–protein interactions controlling SOD1 partitioning are sensitive to environmental conditions. They observed that in dilute solutions (buffer) at high temperature (43 °C), which causes unfolding, SOD1 is sequestered into FUS, and that using Ficoll 70 as a crowder significantly increases SOD1's PC into FUS even at room temperature. Conversely, replacing the crowder with BSA greatly reduces this partitioning, even after 60 minutes of heat stress. Their simulations \cite{Samanta2021} using the implicit solvent force field OPEP showed that the interaction energies of SOD1 with FUS and BSA are similar, differing by only a few kilojoules per mole. As a result, they argued that SOD1 does not have a strong preference for either FUS or BSA, which explains the reduced PC in FUS condensates, consistent with experimental findings. Consequently, the very different behaviors with Ficoll 70 (a PC enhancer) and BSA (a PC reducer) are mainly attributed to the excluded-volume effect of the former and the competing interactions of the latter.



Our results, which include explicit solvent modeling, provide a new perspective that clarifies the origin of these different effects as a consequence of the hydration properties of the CWDs and proteins.
We find that SOD1 has markedly different hydration-dependent landscapes for the free energy with BSA and FUS. In the BSA solution, SOD1 can adopt three typical associative states: SOD1-BSA, SOD1-SOD1, and free SOD1 in the bulk. Conversely, in the FUS solution, SOD1 consistently interacts with FUS, remaining in the same associative state. 
As discussed in Sec. \ref{explicit}, differentiating these states is challenging without water-dependent coordinates.

We argue that the difference in SOD1 behavior arises from the solvent composition: in the BSA solution, the association of SOD1 with another protein leads to a significant increase in bulk water entropy, while in the FUS solution, most water molecules hydrate the proteins without the possibility of entropy gain through water release into the bulk. 
Additionally, the SOD1-SOD1 interaction is slightly more enthalpically favorable than SOD1-BSA binding, further encouraging SOD1 to escape the crowded environment by forming homotypic interactions. 
Since the implicit solvent SOD1-BSA and SOD1-FUS energies \cite{Samanta2021} and water enthalpies (Table~\ref{table:sod1_enthalpy_resutls}) are comparable, we conclude that the difference in solvation free energy—favoring SOD1 in BSA solution—mainly stems from: a) the large entropy gain associated with bulk water in the BSA system, and b) the contribution of homotypic SOD1 interactions. 

The experimental evidence \cite{Samanta2021} indicates that under dilute buffer conditions, SOD1 at high temperature becomes sequestered within FUS condensates. This suggests that the enthalpy contribution of SOD1-BSA plays a crucial role in the observed decrease in PC. 

Additionally, the combined effects of the SOD1-CWD enthalpy change and water entropy contributions do not apply to highly hydrophilic Ficoll 70 polysaccharides, which do not influence the proteins' hydration state. Unlike BSA, Ficoll 70's primary function is to reduce the available volume for SOD1, thereby increasing its local concentration and promoting its sequestration into FUS condensates. 

Furthermore, our analysis enables the evaluation of transition kinetics among different states of SOD1 in BSA solution. On average, SOD1 molecules undergo approximately 90 transitions per microsecond between the three distinct associative states. Additionally, the accessible residence time, $t_R$—the duration a SOD1 molecule remains in a single state—varies over at least three orders of magnitude, ranging from 2~ns to 200~ns. 
We remark, as observed in Sec. \ref{simulation_method}, that our method does not re-weight the OPEP configurations and their dynamics, but instead uses them as a starting point for water equilibration trajectories, which enables us to define water-dependent coordinates.

Two important points should be considered. First, while shorter and longer residence times are possible, our measurements are limited by the temporal resolution (2 ns) and the total simulation duration (1 $\mu$s). Second, the kinetic analysis may be biased due to the way we defined the borders between regions A, B, and C in the free energy landscape. Although these definitions are reasonable—based on the hydration-shell distance related to the associative state—alternative boundary definitions in other free-energy projections could alter the evaluation of transition frequencies and residence times.

\section{Conclusions}
\label{OPEP:conclusions_SOD1}

Recent research increasingly highlights the vital role of solvents in the formation and regulation of biomolecular condensates \cite{Watson:2023aa}, supported by compelling experimental evidence on hydration dynamics \cite{Konig:2024aa} and valuable mechanistic insights from simulations \cite{SCHAFER2025103026}.
Given this growing body of evidence, developing a combined computational and theoretical approach to quantify the role of hydration in protein sequestration is both timely and essential. Our current work outlines a method aimed at bridging the gap between molecular-level solvent effects and mesoscale condensate properties, thereby offering a more comprehensive understanding of the underlying mechanisms.

We investigate the impact of the environment on the behavior of SOD1 proteins within highly concentrated solutions of BSA and FUS. The choice of these co-solutes is motivated by their distinct conformational properties—BSA being globular and FUS intrinsically disordered—and aims to rationalize {\it in vitro} observations showing a decrease in the partition coefficient of SOD1 in FUS condensates when BSA proteins serve as cytomimetic crowders, as an alternative to synthetic Ficoll 70 \cite{Samanta2021}. Our methodology (Sec. \ref{couple_cvf_opep}) maps trajectories obtained using the OPEP implicit solvent model \cite{SterponeReview2014, Samanta2021} into a discretized OPEP-dis configuration. The OPEP-dis representation can then be hydrated with CVF water and equilibrated along new MC trajectories, each corresponding to an original OPEP configuration. This approach does not change the free energy of the system because the OPEP implicitly accounts for water contributions. However, it allows us to calculate the water-water and water-protein interactions for the enthalpy along the system's trajectory, as well as to project the Gibbs free energy along water-dependent generalized coordinates.

The calculation of water contribution to the enthalpy shows an overall difference of approximately 0.6 kJ/mol, favoring SOD1 in FUS solution. This difference is likely due to the increased number of water-protein contacts in the FUS solution compared to BSA. Although this varies depending on the parameters used, it does not explain the interaction energy difference favoring BSA observed in the implicit-solvent analysis of Ref. \cite{Samanta2021}. Consequently, it cannot account for the mechanism limiting SOD1 partitioning into FUS in the presence of BSA crowders. Instead, this mechanism is revealed by the explicit-solvent free energy landscape analysis, which offers more detailed and complementary insights than the implicit-solvent energy calculations \cite{Samanta2021}. 

This is especially clear in the BSA case. When the free energy is projected onto implicit solvent generalized coordinates, the landscape displays a flat basin encompassing both non-interacting and interacting proteins at various distances. Projecting onto a solvent-dependent generalized coordinate that measures hydration state reveals that SOD1 proteins can explore three different associative states: (A) SOD1-BSA, (B) SOD1-SOD1, and (C) free SOD1 in the bulk water. Our results indicate that these three states are interconnected, with state A serving as an intermediate between B and C. A free energy barrier of a few $k_BT$ separates states B and C, which can be overcome by thermal fluctuations. Additionally, we observe that states A and B are more frequently visited than state C, with a slight preference for B. Residence times in each state can reach up to 200 ns; however, the transition rates among states can be as high as 90 $\mu$s$^{-1}$, with more frequent transitions between A and B and fewer between B and C, due to the free energy barrier that separates them. 
The dominance of states A and B is a consequence of the significant entropy increase resulting from the release of hydration water into the bulk phase during protein association. Additionally, the homotypic interactions of SOD1 further favor state B. 

This water-entropy gain mechanism is not significant for SOD1 within FUS biocondensates. In these systems, FUS chains are evenly distributed throughout, resulting in most water molecules hydrating either FUS or SOD1 amino acids. Consequently, no water can be released into the bulk. Nonetheless, the hydration generalized coordinate still provides valuable insight into the system's free energy landscape.
In the implicit-solvent projection, the free energy landscape appears to feature a weakly-interacting SOD1-FUS state separated by approximately $5 k_BT$ from a metastable, strongly-interacting state. Conversely, the projection onto the explicit solvent coordinate reveals a single basin characterized by SOD1's varying hydration levels interacting with FUS.

In conclusion, although enthalpy calculations rely on parameter optimization, our hydration method enables us to project free energy onto solvent generalized coordinates. This clarifies the essential role water plays in shaping the entropic mechanisms that influence the reduced partition coefficient of SOD1 within FUS biomolecular condensates in the presence of BSA crowders. 

This mechanism does not affect highly hydrophilic Ficoll 70, which mainly reduces the volume for SOD1 and promotes its inclusion in FUS condensates, unlike BSA proteins. Overall, our approach offers new insights into previous implicit-solvent simulations and opens avenues for large-scale, biologically relevant simulations that explicitly incorporate water's contribution.

From a broader perspective, if LBMD can incorporate dynamic hydration effects through hydrodynamic interactions, disentangling pairwise and many-body thermodynamic HB solvation contributions with minimal computational cost during simulations could be a major breakthrough. This would open up exciting possibilities for mesoscale simulations of biomolecular processes. Such integration would allow for more realistic and efficient modeling of protein sequestration and phase behavior, effectively bridging the gap between microscopic hydration phenomena and mesoscale condensate dynamics.




\section*{Acknowledgments}
We thank Emeline Laborie for providing data related to the OPEP model.
This work was supported by the Spanish Ministerio de Ciencia e Innovaci\'on/Agencia Estatal de Investigaci\'on [grant number MCIN/AEI/ 10.13039/ 501100011033]; the European Commission “ERDF A way of making Europe” [grant number PID2021-124297NB-C31]; the Universitat de Barcelona [grant number 5757200 APIF\_18\_19].
G.F. acknowledges the support from the Ministry of Universities 2023-2024 Mobility Subprogram within the Talent and its Employability Promotion State Program (PEICTI 2021-2023) and the Visitor Program of the Max Planck Institute for The Physics of Complex Systems for supporting a visit started in November 2022.

\bibliography{ref-CVF}

\clearpage

\appendix

\section{Selection of CVF parameters at the hydration shell}
\label{si:cvf_parameters}

At variance with the bulk CVF parameters, that are set to reproduce the experimental equation of state of water around ambient conditions~\cite{CoronasJMolLiq2025}, the calibration of the CVF parameters at the hydration shell is still missing. As a first approximation, we adapt a set of parameters based on a previous work of hydrated proteins in a water monolayer~\cite{BiancoPRX2017} to the CVF bulk case. The hydrophobic interface strengthens the water-water hydrogen bonding in the first hydration layer~\cite{Petersen2009,Tarasevich2011,Sarupria:2009ly}, therefore we assume $J^{\rm PHO}_1 > J$ and $J^{\rm PHO}_{\sigma, 1}>J_\sigma$. Moreover, the local density of water at the hydrophobic interface increases upon pressurization~\cite{Sarupria:2009ly,Das2012,Ghosh2001,Dias:2014aa}. The CVF model reproduces this behavior assuming that the local volume change associated with the formation of a HB at the first hydration shell depends linearly on $P$ as $v_{{\rm HB},1}^{\rm PHO} \equiv (1-k\cdot P) v_{{\rm HB},1}^{\rm PHO}$, where $v_{{\rm HB},1}^{\rm PHO}=v_{\rm HB}$ is the volume change at zero pressure (internal CVF units), and $k>0$. We note that the variation of $v_{{\rm HB},1}^{\rm PHO}$ on $P$ is not relevant for our purposes, as this work is performed entirely at ambient thermodynamic conditions $T=300$~K, $P=1$~atm. However, we apply this formula to estimate the corresponding $v_{{\rm HB},1}^{\rm PHO}$ at $P=1{\rm~atm}$. Following Ref.~\cite{BiancoPRX2017}, we assume $J^{\rm PHO, 1} = 4J$, $J^{\rm PHO, 1}_\sigma = 4J_\sigma$, and $v_{{\rm HB},1}^{\rm PHO}(1{\rm ~atm}) = 4v_{\rm HB}$. Regarding the hydrophilic hydration shell, Ref.~\cite{BiancoPRX2017} assumes that the HBs are not affected by the interface, so $J^{\rm PHI,1}=J$, $J^{\rm PHI,1}_\sigma=J$, and $v_{{\rm HB},1}=v_{\rm HB}$ (see Table~\ref{table:CVFExtended_parameters_BiancoPRX}). 

We set the values of the second and third layers equal to the values of the first layer. An alterantive choice could be a smooth transition bewteen the values from the parameters at the first layer to the bulk, and this is to be explored in future refinement of the model. Finally, for the MIX hydration layer, we set $J^{\rm MIX}_l\equiv(J^{\rm PHO}_l+J^{\rm PHI}_l)/2$, $J^{\rm MIX}_{\sigma,l}\equiv(J^{\rm PHO,l}_\sigma+J^{\rm PHI,l}_\sigma)/2$, and $v_{{\rm HB},l}^{\rm MIX}\equiv(v_{{\rm HB},l}^{\rm PHO}+v_{{\rm HB},l}^{\rm PHI})/2$.

To explore how the results are affected by considering HBs at the hydrophilic interface different from those at the bulk, we tested a second set of parameters in \cite{CoronasThesis}. The second set assumes that the enthalpy change due to HB formation is $1.8$ times stronger at the PHO interface than the bulk, and $1.5$ times stronger at the PHI interface. However, the results show no qualitative difference and are not included in this work.

For $S_i^{\rm w}$, we set $S_i^{\rm w}<0$ if the residue is hydrophilic, and $S_i^{\rm w}=0$, otherwise. In the OPEP-dis representation, large residues tend to occupy more cells than small residues. Thus, the number of water-residue contacts will be larger for bigger residues. To balance the energy contribution of large and small residues, we set $S_i^{\rm w}=-0.1$ if the radius of the bead is larger that the water van der Waals diameter $r_k\geq r_0$, and $S_i^{\rm w}=-0.5$ oterwise, see Table~\ref{table:amino_acids}.

\begin{table*}
\centering
\begin{tabular}{| c | c | c | c | c |}
\hline
 & BULK & PHO & PHI & MIX \\
\hline
$J/4\epsilon$ & 0.5 & 2 & 0.5 & 1.25 \\
\hline
$J_\sigma/4\epsilon$ & 0.08 & 0.32 & 0.08 & 0.2  \\
\hline
$v_{\rm HB}/v_0$ & 0.6 & 4 & 0.6 & 2.3\\
\hline
$k \cdot (4\epsilon/v_0)$ & 0 & 0.889 & 0 & 0.773 \\
\hline
$J^{\rm eff}(P=1{\rm~atm})/4\epsilon$ & 0.23 & 0.92 & 0.23 & 0.575 \\
\hline
$v_{\rm HB}(P=1{\rm~atm})/v_0$ & 0.6 & 2.4 & 0.6 & 1.5 \\
\hline 
\end{tabular}
\caption{Set of parameters describing HBs energy and volume at the hydration shell. The values reported for BULK have been parametrized to reproduce the experimental density and enthalpy of water, and fluctuations, around ambient conditions~\cite{CoronasJMolLiq2025}. The parameters for PHOB, PHIL and MIX follow Ref.~\cite{BiancoPRX2017}, and are equal for the three hydration layers. The last two rows show the resulting enthalpy and volume change due to HB formation at ambient pressure: $J^{\rm eff} \equiv J-Pv_{\rm HB}$ and $v_{\rm HB} \equiv (1-k\cdot P)v_{\rm HB}$ respectively, with $P=0.45 (4\epsilon/v_0)= 0.1{\rm~MPa}$~\cite{CoronasJMolLiq2025}.}
\label{table:CVFExtended_parameters_BiancoPRX}
\end{table*}

\begin{table*}
\centering
\begin{tabular}{ |c|c|c|}
\cline{0-2}
 Amino acid & vdW radius [\AA] Ref.~\cite{creighton1993} & $S^{\rm w} [4\epsilon]$ \\
\cline{0-2}
ARG & 3.3  &  -0.1  \\
\cline{0-2}
LYS & 3.2  &   -0.1   \\
\cline{0-2}
ASP &  2.8 &  -0.5  \\
\cline{0-2}
GLU & 3.0 &  -0.1  \\
\cline{0-2}
ASN & 2.8  &  -0.5  \\
\cline{0-2}
GLN &  3.0  & -0.1   \\
\cline{0-2}
CYS &  2.7 & -0.5   \\
\cline{0-2}
MET & 3.1  &  0   \\
\cline{0-2}
HIS & 3.0   & -0.1  \\
\cline{0-2}
SER & 2.6  &  -0.5  \\
\cline{0-2}
THR & 2.8  &  -0.5  \\
\cline{0-2}
VAL & 2.9  &  0    \\
\cline{0-2}
LEU & 3.1  &   0 \\
\cline{0-2}
ILE & 3.1  &  0  \\
\cline{0-2}
PHE & 3.2  &  0  \\
\cline{0-2}
TYR & 3.2  &  -0.1  \\
\cline{0-2}
TRP & 3.4  &  0    \\
\cline{0-2}
GLY & 2.3  & -0.5   \\
\cline{0-2}
ALA &  2.5  &  0 \\
\cline{0-2}
PRO & 2.8  &  0  \\
\cline{0-2}
\end{tabular}
\caption{Table of amino acid-water interactions. In the second column, we report the van der Waals (vdW) radius of each amino acid. In the third column, we report the interaction energy per contact with BF water. If the residue is hydrophobic, then $S^{\rm w}=0$. Otherwise, $S^{\rm w}/4\epsilon=-0.5$ if the radius of the amino acid is $r_i<r_0=2.9$~\AA~and $S^{\rm w}/4\epsilon=-0.1$ if $r_0<r_i<2r_0=5.8$~\AA, with $\epsilon \equiv 5.5 {\rm~kJ/mol}$~\cite{Henry2002,CoronasBook2022}.}
\label{table:amino_acids}
\end{table*}

\begin{table*}[h]
\centering
\begin{tabular}{| c | c | c | c | c |}
\hline
\multicolumn{4}{|c|}{ENTHALPY [kJ/mol] }\\
\hline
  & SOD1-BSA & SOD1-FUS & $\Delta$ [FUS-BSA]    \\
\hline
Total: $H \equiv H^{\rm bulk} + H^{\rm hyd} $
 & -14.0248(9) & -14.6546(14) & -0.630(2) \\
\hline
$H^{\rm hyd} \equiv H^{\rm hyd}_{\rm SOD1} + H^{\rm hyd}_{\rm CWD} + H^{\rm hyd}_{\rm mix}$ & -1.922(2) & -3.933(2) & -2.011(4)  \\
\hline
$H^{\rm hyd}_{\rm SOD1} \equiv H^{\rm w,w}_{\rm SOD1} + E^{R,w}_{\rm SOD1} $  &  -0.2816(8) & -0.1958(9) & +0.084(2)  \\
\hline
$H^{\rm hyd}_{\rm CWD} \equiv H^{\rm w,w}_{\rm CWD} + E^{R,w}_{\rm CWD} $ & -1.6298(14) & -3.689(3) & -2.059(4) \\
\hline
$H^{\rm hyd}_{\rm mix} \equiv H^{\rm w,w}_{\rm mix} $ & -0.0107(2) & -0.0478(2) & -0.0371(4) \\
\hline
\end{tabular}
\caption{CVF water enthalpy calculations expressed in kJ/mol. First line: total enthalpy of water, including isotropic van der Waals, water-water HBs and water-residue interactions. Second line: enthalpy of water within the hydration shell, including water-water HBs and water-residue interactions. Third-to-fifth lines: enthalpy of water within the hydration shell separated into SOD1, crowder (FUS/BSA) and mixed contributions. The fourth column corresponds to the difference between the third and the second columns. Results calculated considering the set of parameters in Table~\ref{table:CVFExtended_parameters_BiancoPRX}.} 
\label{table:sod1_enthalpy_resutls}
\end{table*}

\begin{figure*}[]
    \centering
    \includegraphics[scale=0.8]{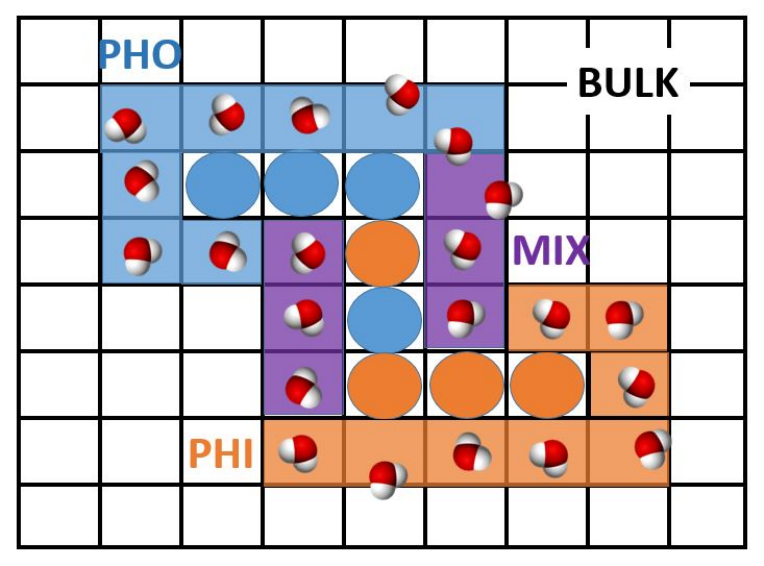}
    \caption{Schematic representation of the hydrataion shell in the CVF model. Circles indicate cells occupied by protein residues; otherwise, it is occupied by water. For the sake of simplicity we show a hydration shell of $n_l=1$ layer. Colored cells indicate the hydration shell. Water molecules in blue cells hydrate hydrophobic residues (blue circles), molecules in orange cells hydrate hydrophilic residues (orange circles), and molecules in violet cells hydrate both hydrophobic and hydrophilic residues (mixed, MIX shell). Subsequent layers $n_l>1$ of the hydration shell are defined and classified in the same manner.}
    \label{fig:hyd_shell_def}
\end{figure*}

\begin{figure*}
    \centering
    \includegraphics[scale=0.4]{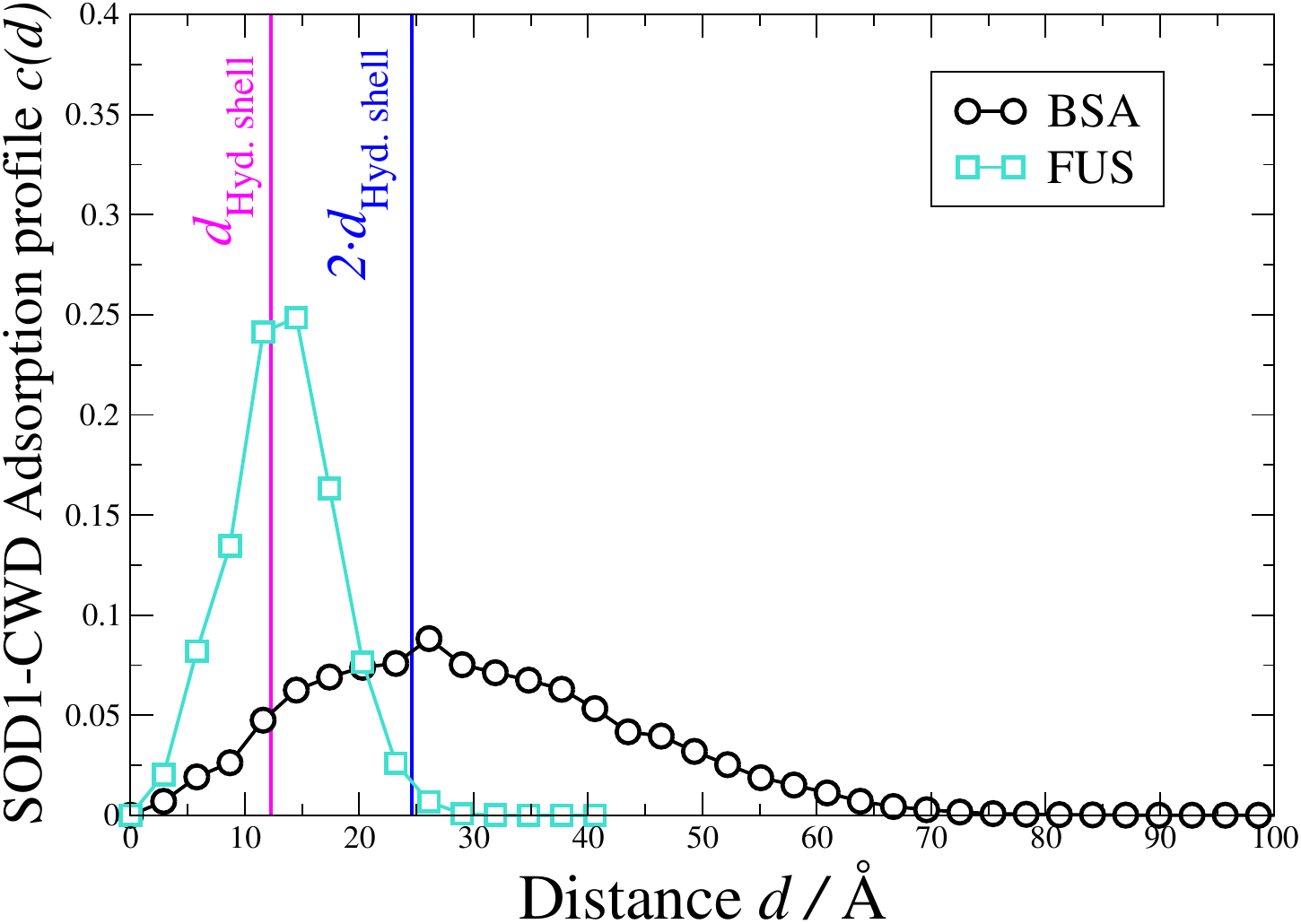}
    \caption{Adsorption profiles $c(d)$ of SOD1 in BSA     (circles) and SOD1 in FUS (squares) solutions, as defined in the text. Magenta and blue lines indicate the distances of     one and two hydration shells, respectively.}
    \label{fig:sod1_adsorption_profile}
\end{figure*}

\begin{figure*}
    \centering
     \includegraphics[scale=0.5]{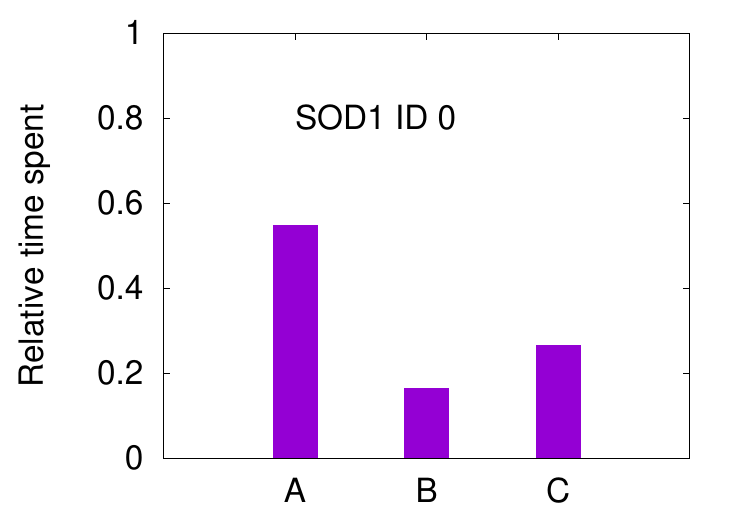}
     \includegraphics[scale=0.5]{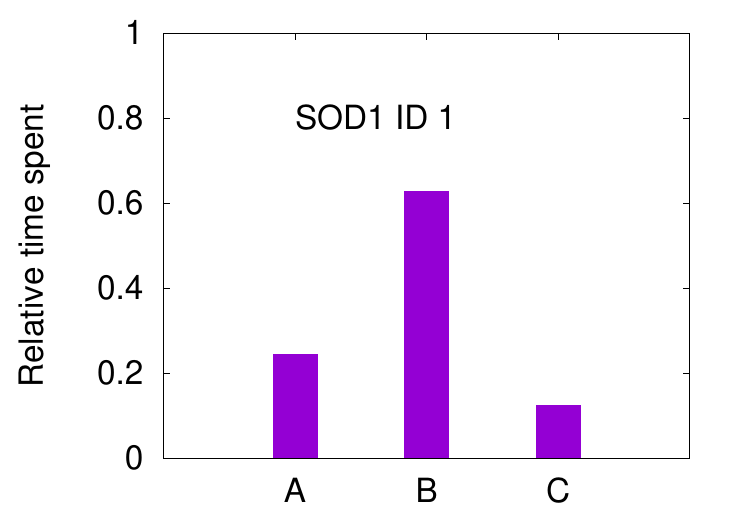}
    \includegraphics[scale=0.5]{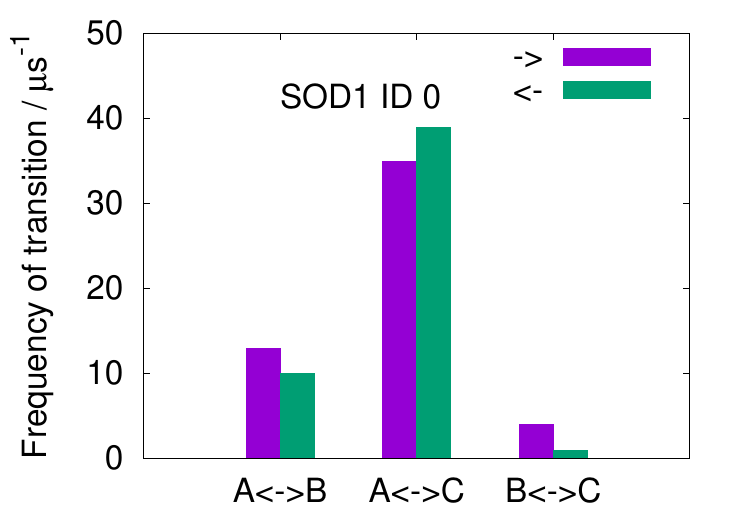}
    \includegraphics[scale=0.5]{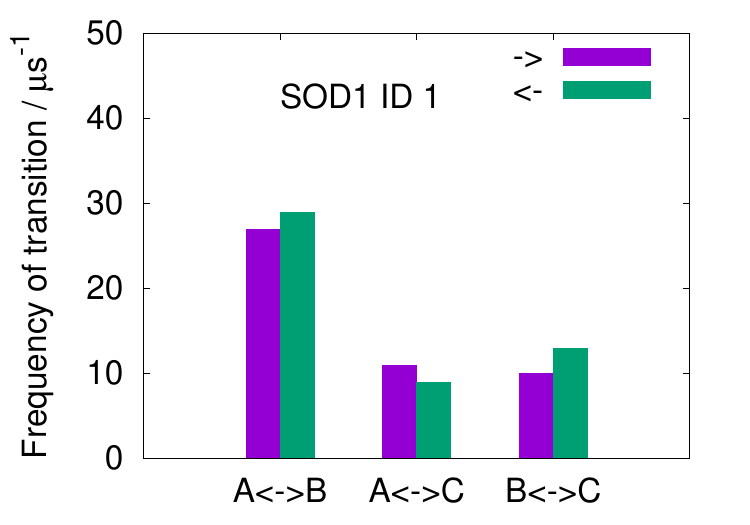}
    \includegraphics[scale=0.43]{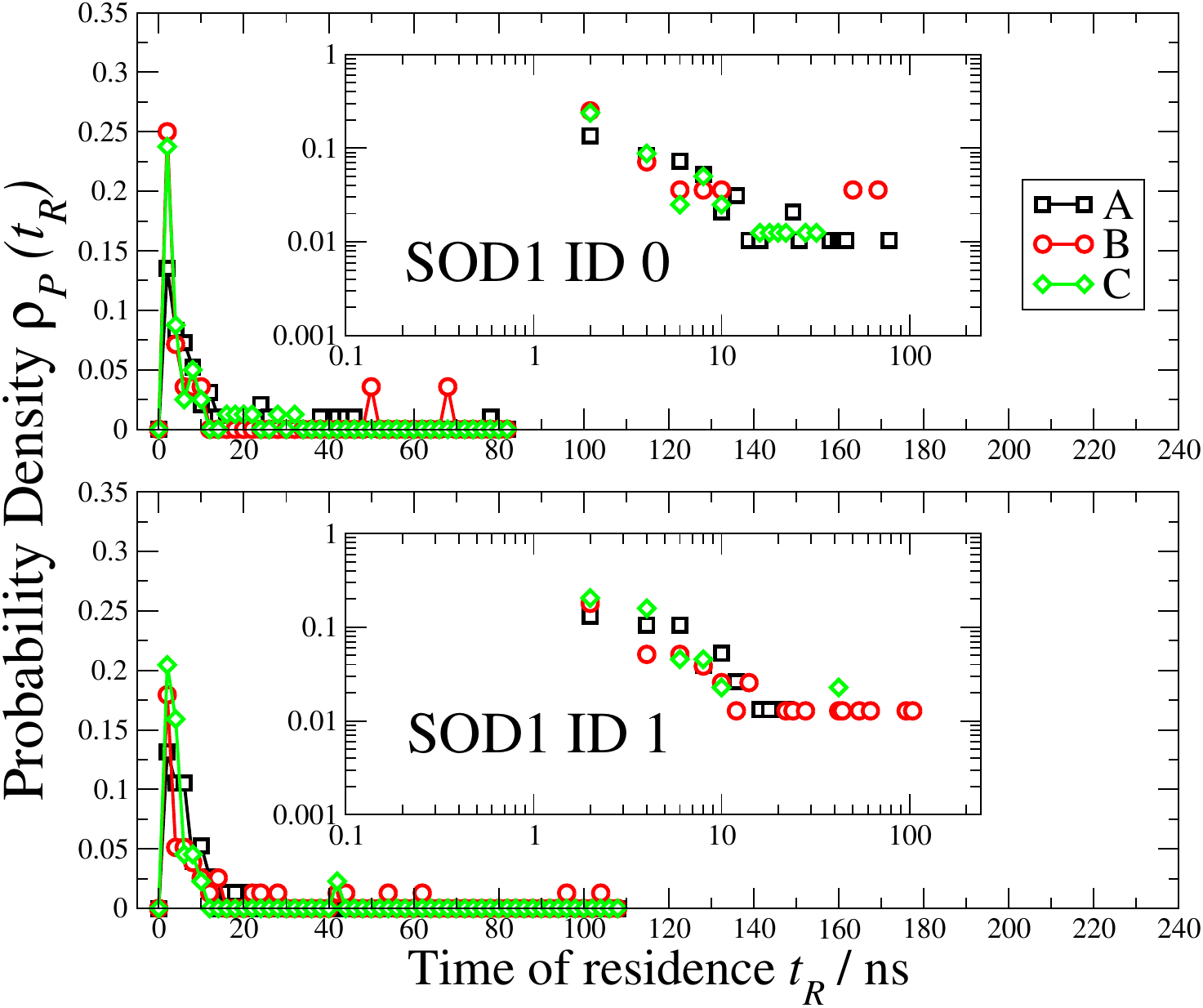}
    \caption{Kinetic analysis of      individual SOD1s. This figure* refers to SOD1-ID 0 and 1, as      labeled in each panel. Top: Time      spent in each state, relative to the total simulation time. Center:      Frequency of transition $\nu_{X\to Y}$ defined as      the number of transitions per      unit time. Bottom: Probability      density of the times of residence      for each configuration A, B or C.      Lines are guides to the eyes connecting the symbols.}
    \label{fig:transition_residence_0_1}
\end{figure*}

\begin{figure*}
    \centering
    \includegraphics[scale=0.5]{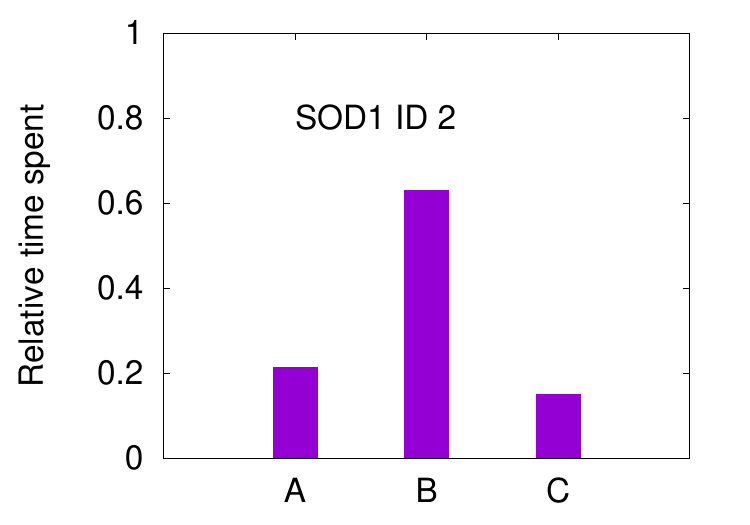}
     \includegraphics[scale=0.5]{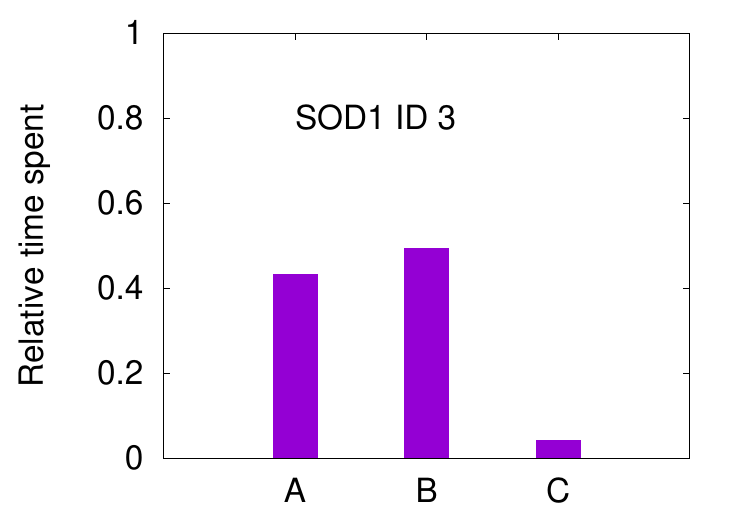}
    \includegraphics[scale=0.5]{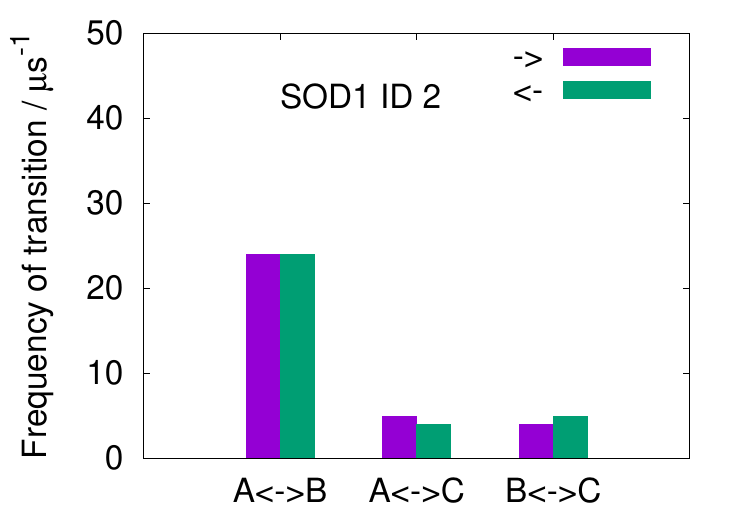}
    \includegraphics[scale=0.5]{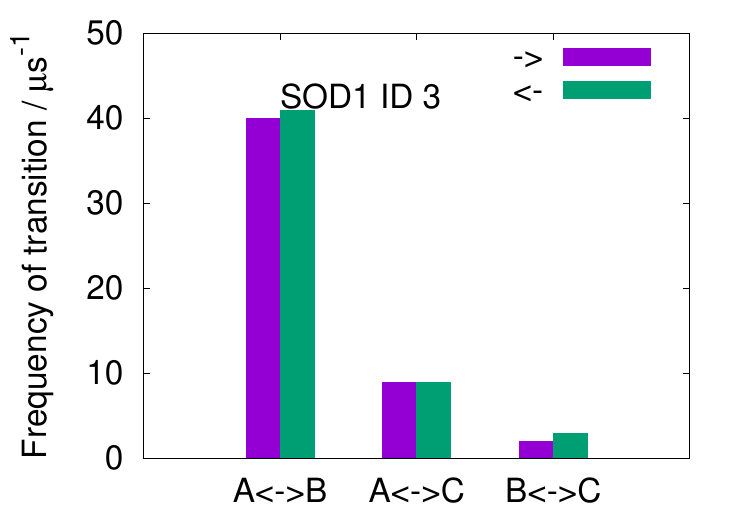}
    \includegraphics[scale=0.43]{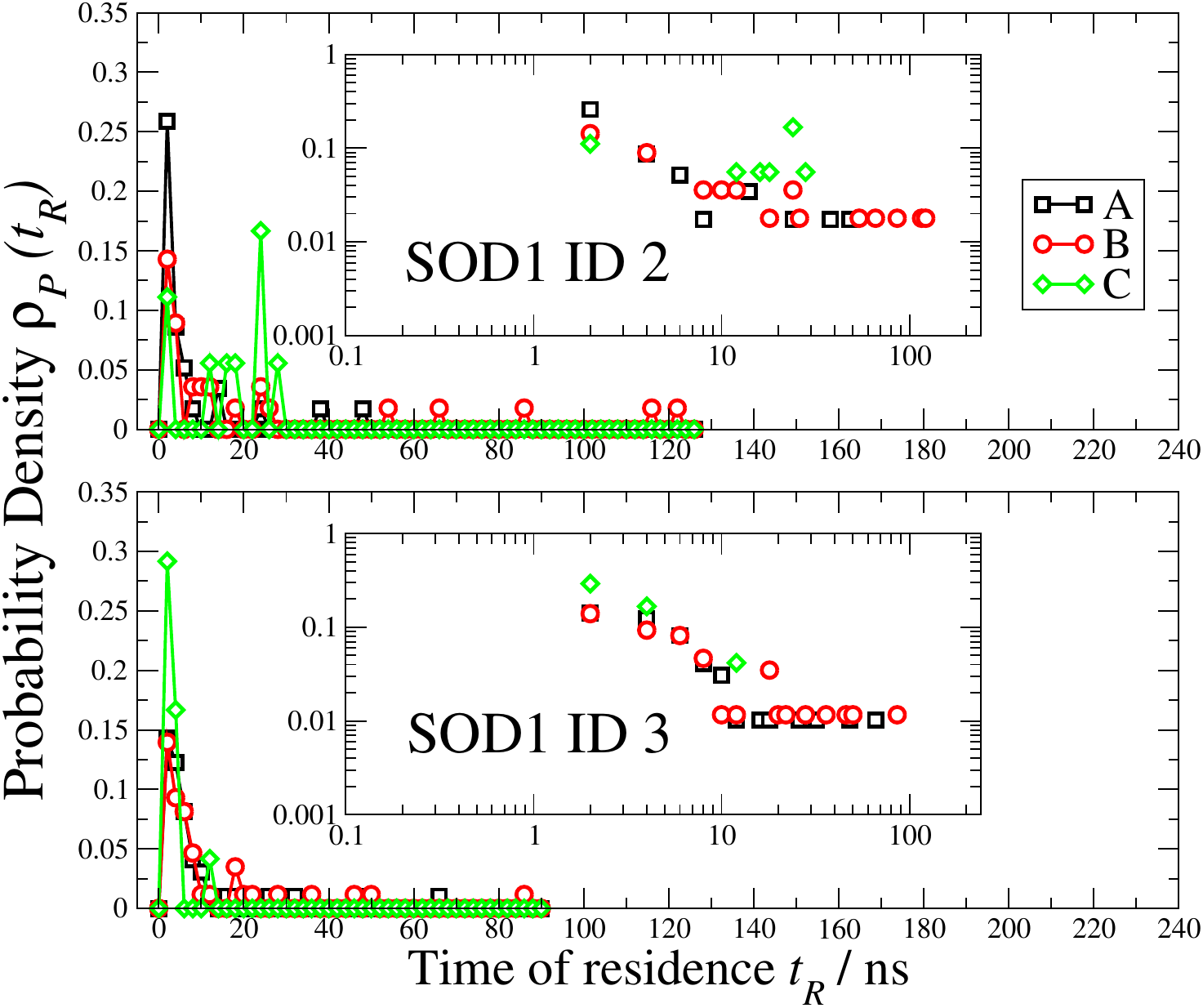}
    \caption{Same as in Fig.~\ref{fig:transition_residence_0_1}, but for SOD1s ID 2 and 3.}
    \label{fig:transition_residence_2_3}
\end{figure*}

\begin{figure*}
    \centering    \includegraphics[scale=0.5]{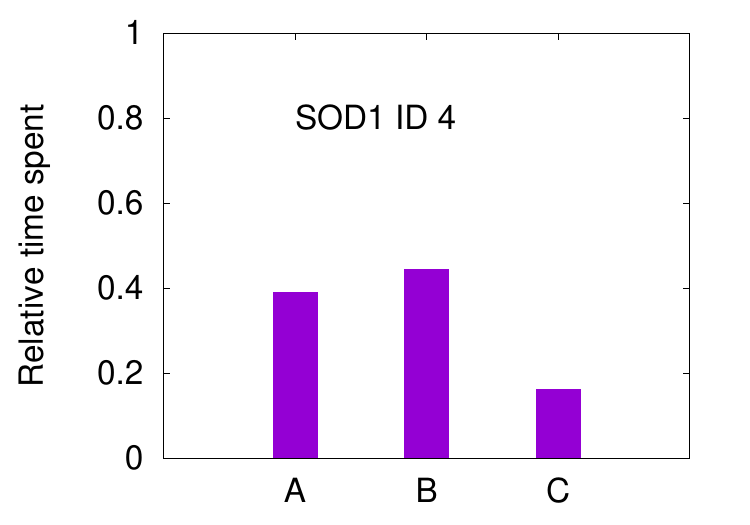}
    \includegraphics[scale=0.5]{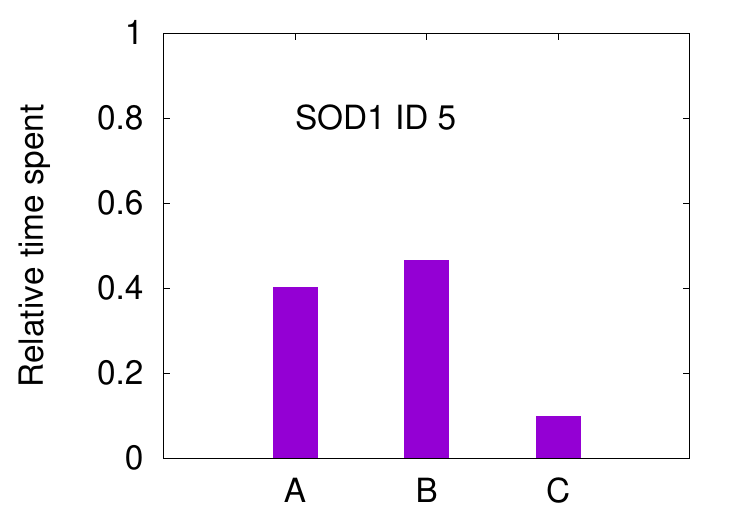}
    \includegraphics[scale=0.5]{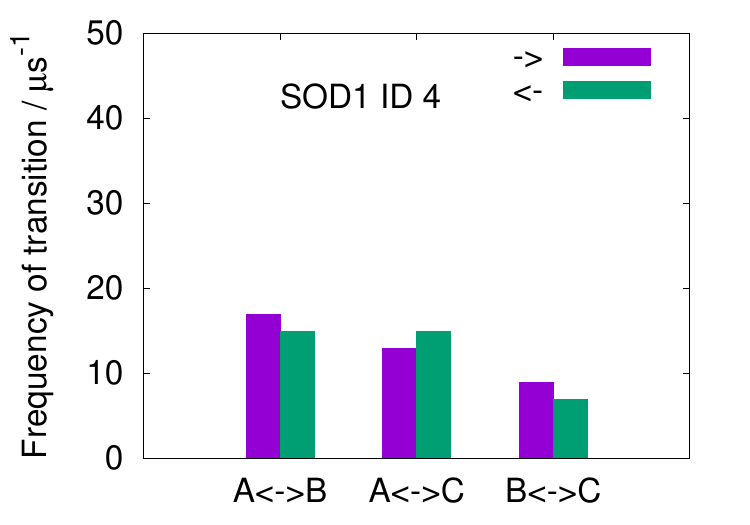}
    \includegraphics[scale=0.5]{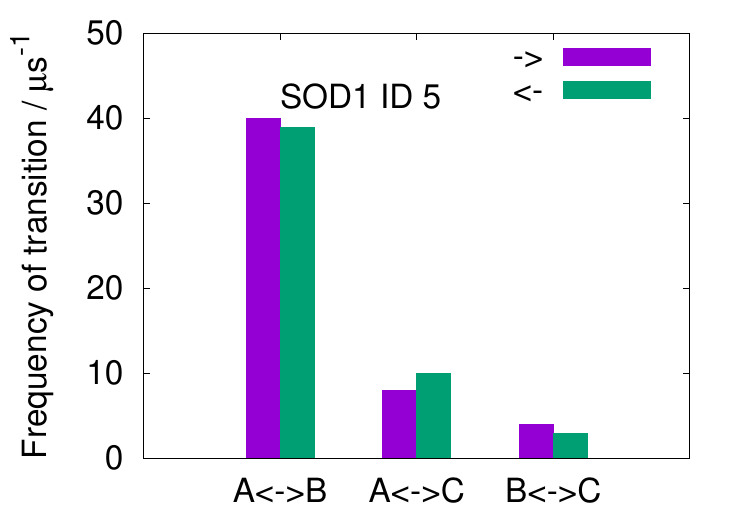}
    \includegraphics[scale=0.43]{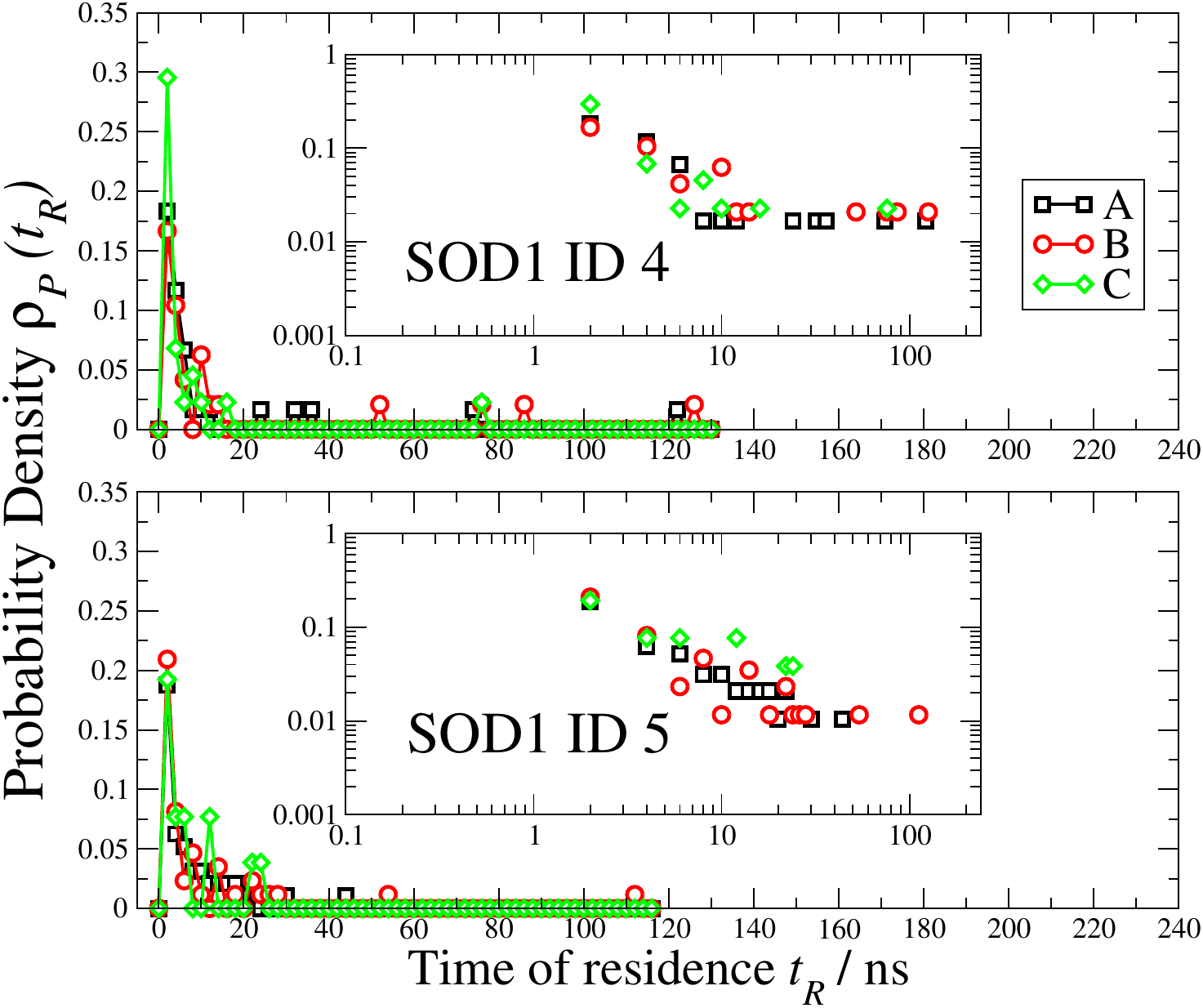}
    \caption{Same as Fig.~\ref{fig:transition_residence_0_1} but for SOD1s ID 4 and 5.}
    \label{fig:transition_residence_4_5}
\end{figure*}

\begin{figure*}
    \centering    \includegraphics[scale=0.5]{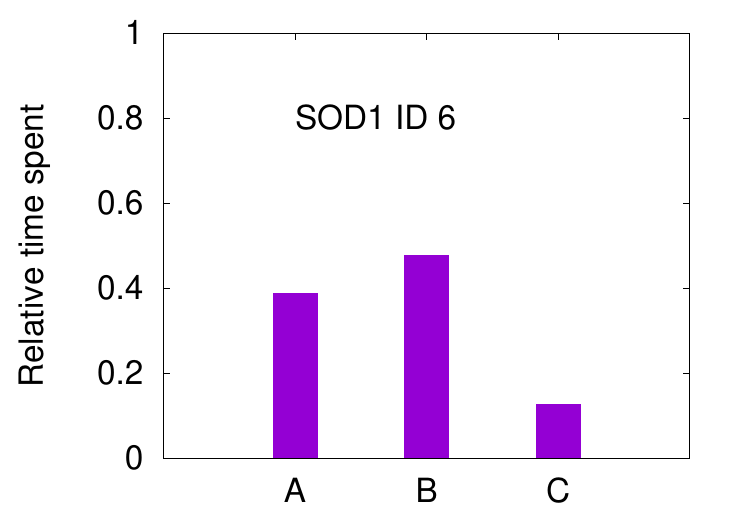}
     \includegraphics[scale=0.5]{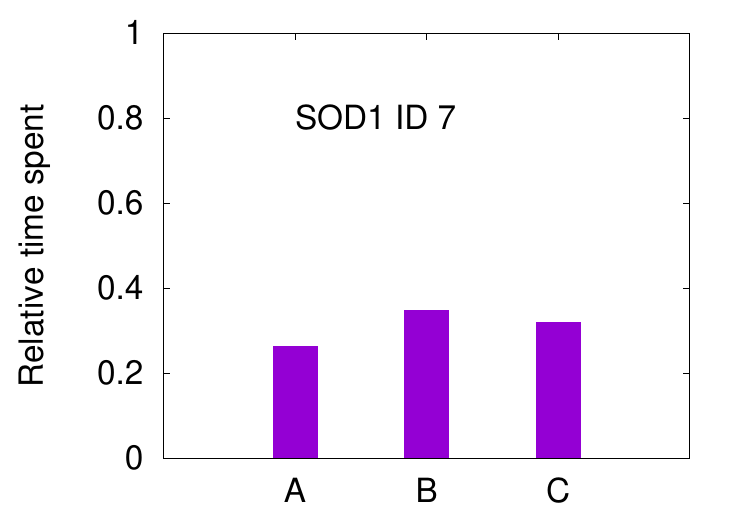}
    \includegraphics[scale=0.5]{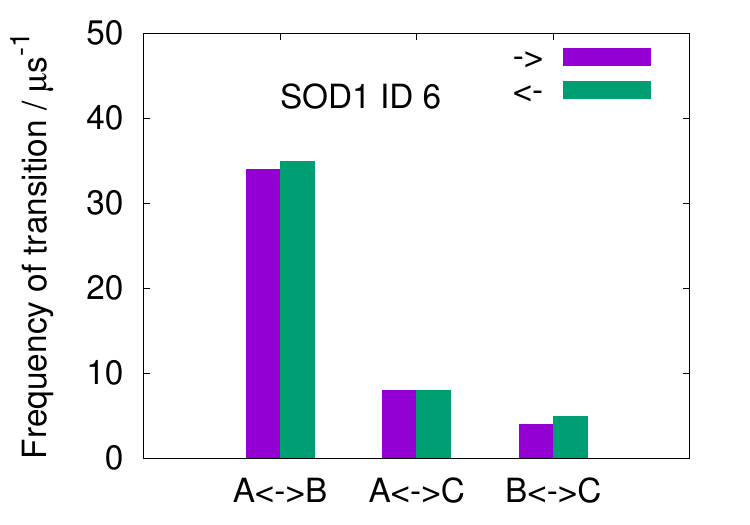}
    \includegraphics[scale=0.5]{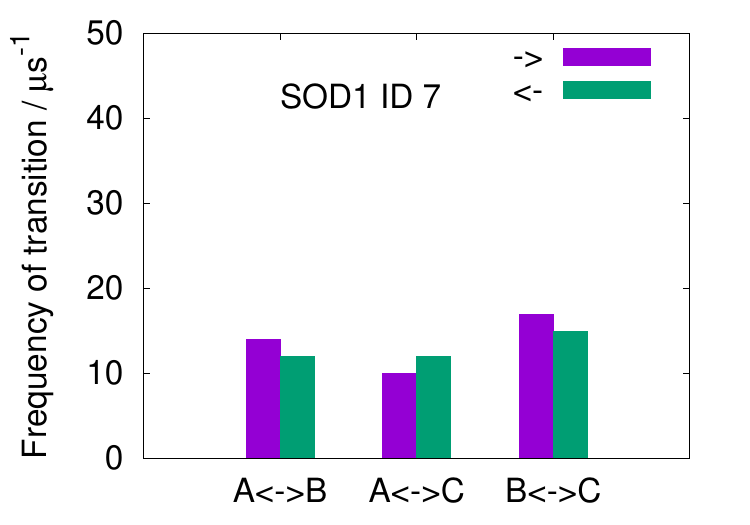}
    \includegraphics[scale=0.43]{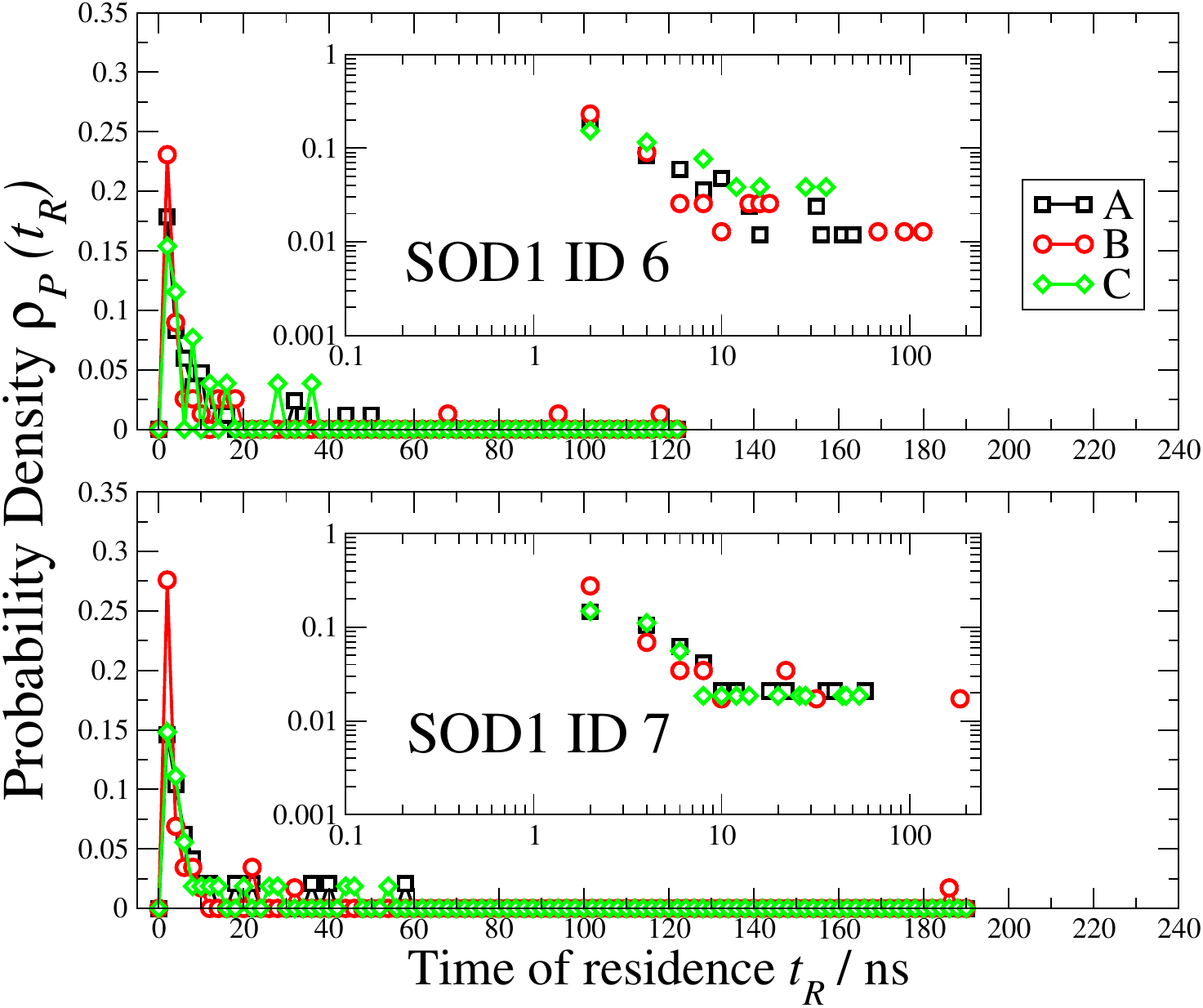}
    \caption{Same as in Fig.~\ref{fig:transition_residence_0_1}, but for SOD1s ID 6 and 7.}
    \label{fig:transition_residence_6_7}
\end{figure*}

\begin{figure*}
    \centering
    \includegraphics[scale=0.5]{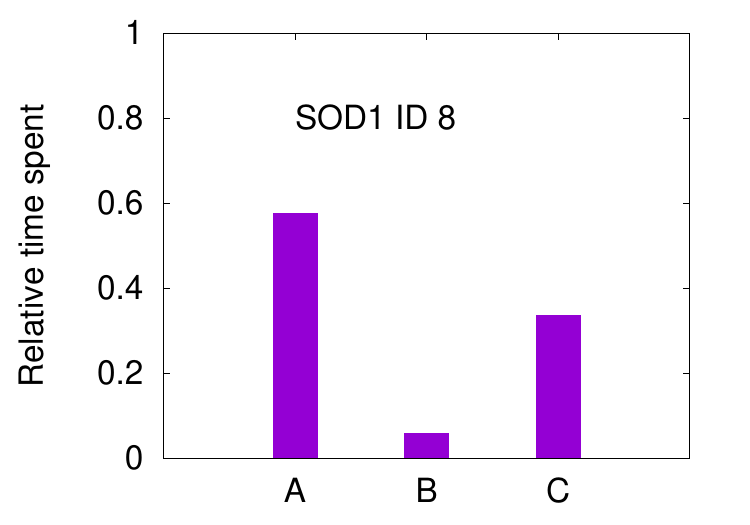}
     \includegraphics[scale=0.5]{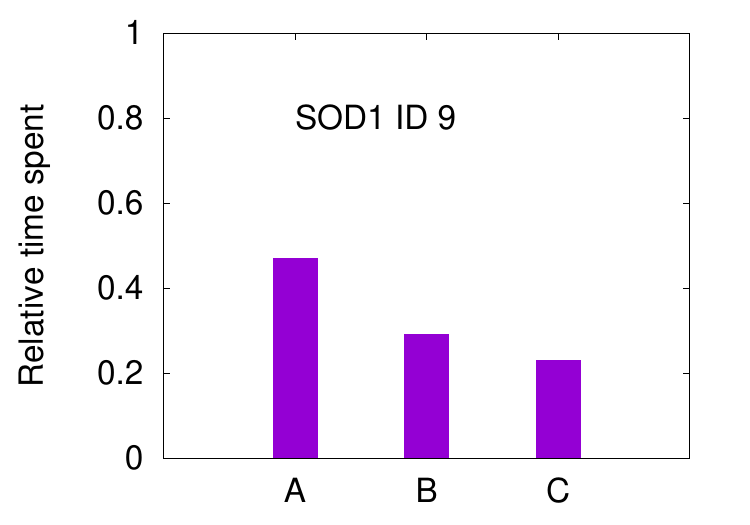}
    \includegraphics[scale=0.5]{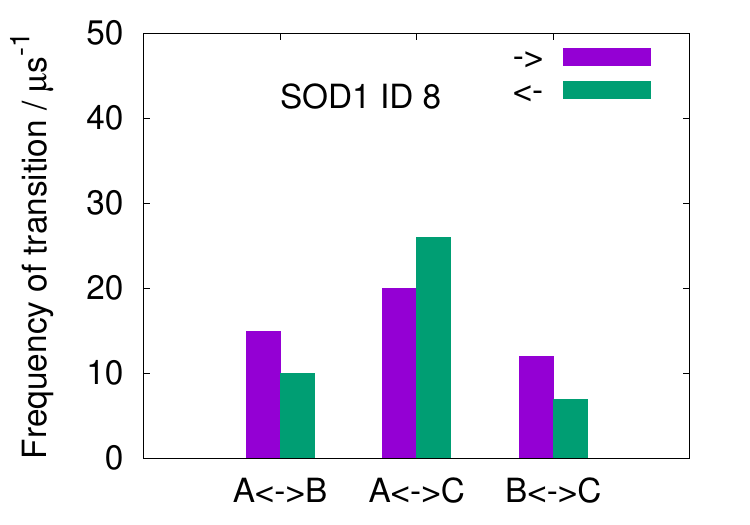}
    \includegraphics[scale=0.5]{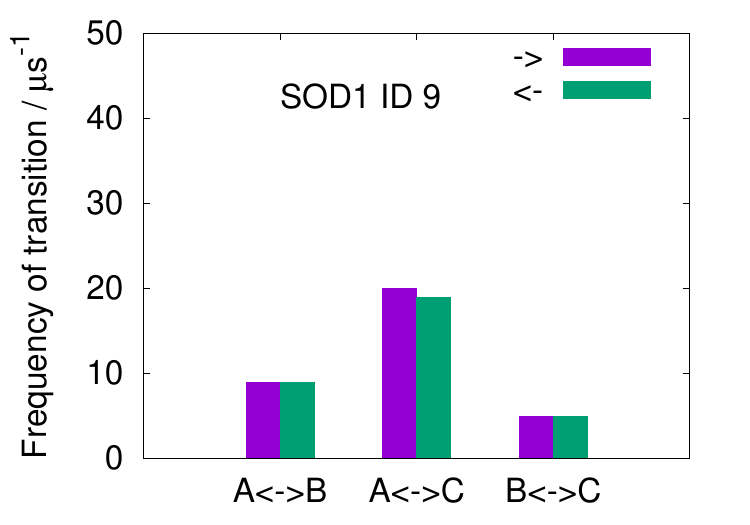}
    \includegraphics[scale=0.43]{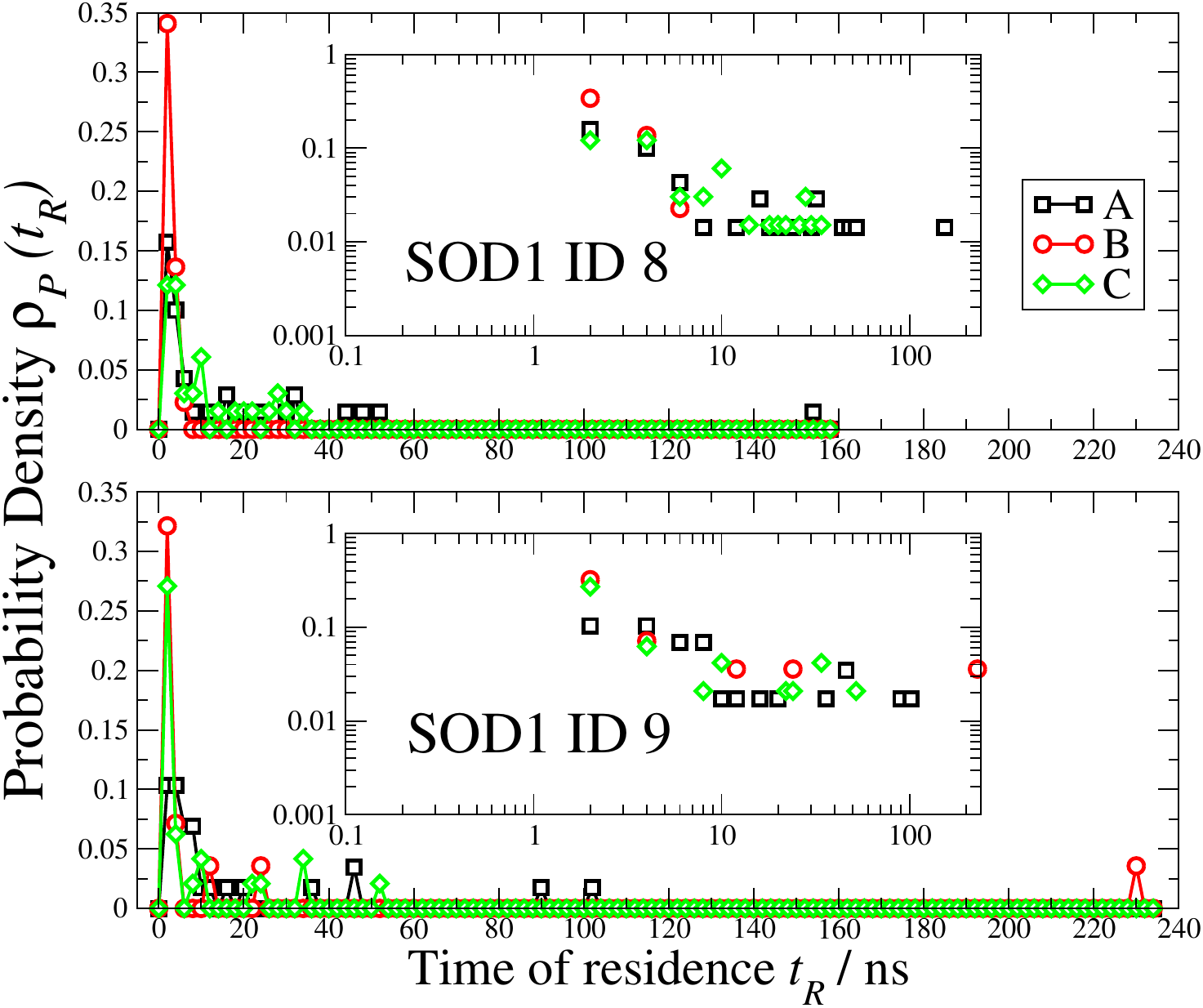}
    \caption{Same as in Fig.~\ref{fig:transition_residence_0_1}, but for SOD1s ID 8 and 9.}
    \label{fig:transition_residence_8_9}
\end{figure*}


\begin{figure*}
    \centering
   SOD1-BSA ADSORP. ${\cal C}[$~\AA$]$ \ \ \  HYD. SHELL $10^3 N_{\rm hyd, SOD1}$  \ \ \ \ \ \  STATE A / B / C \\
    \centering
    \includegraphics[width=0.28\textwidth]{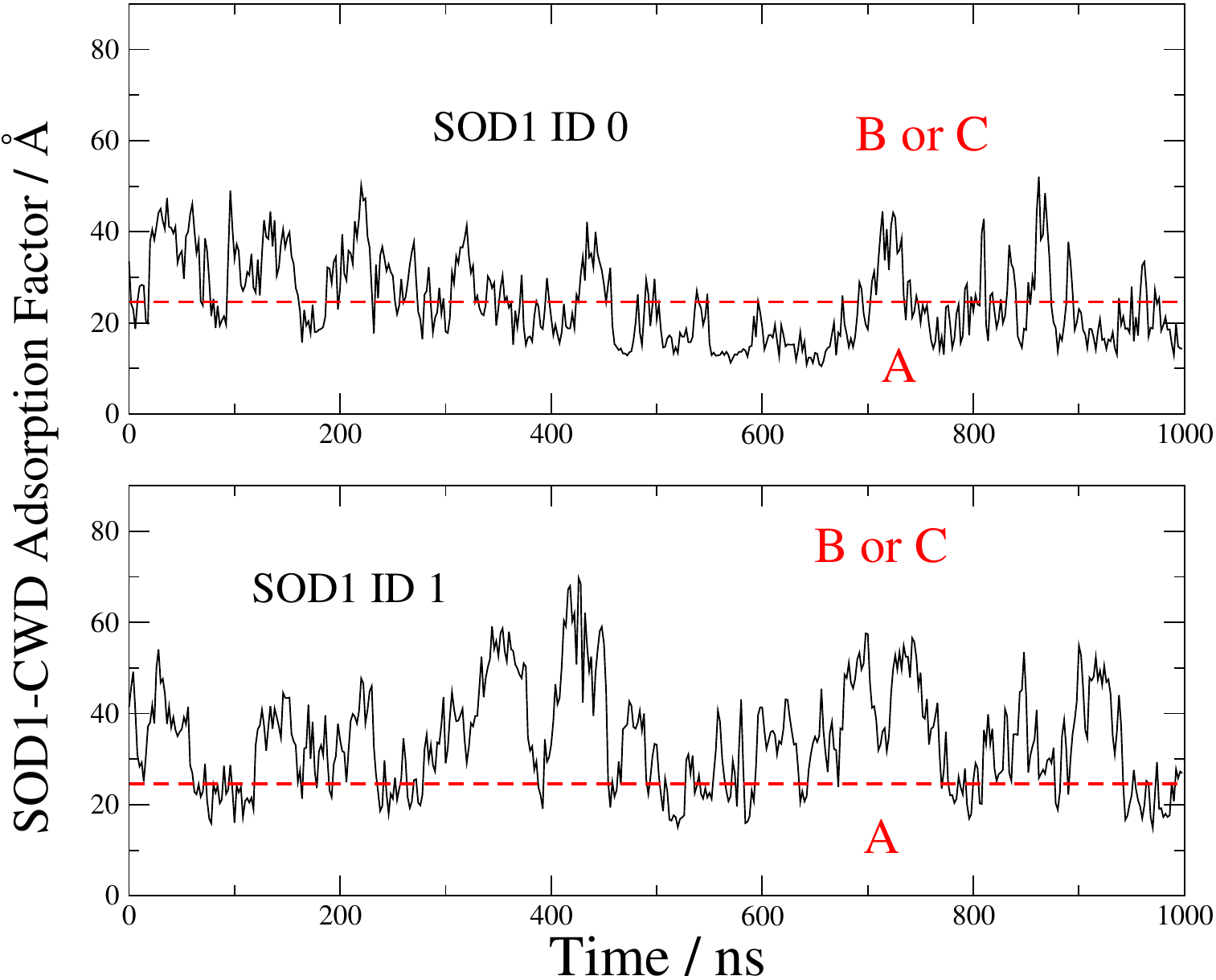}
    \includegraphics[width=0.28\textwidth]{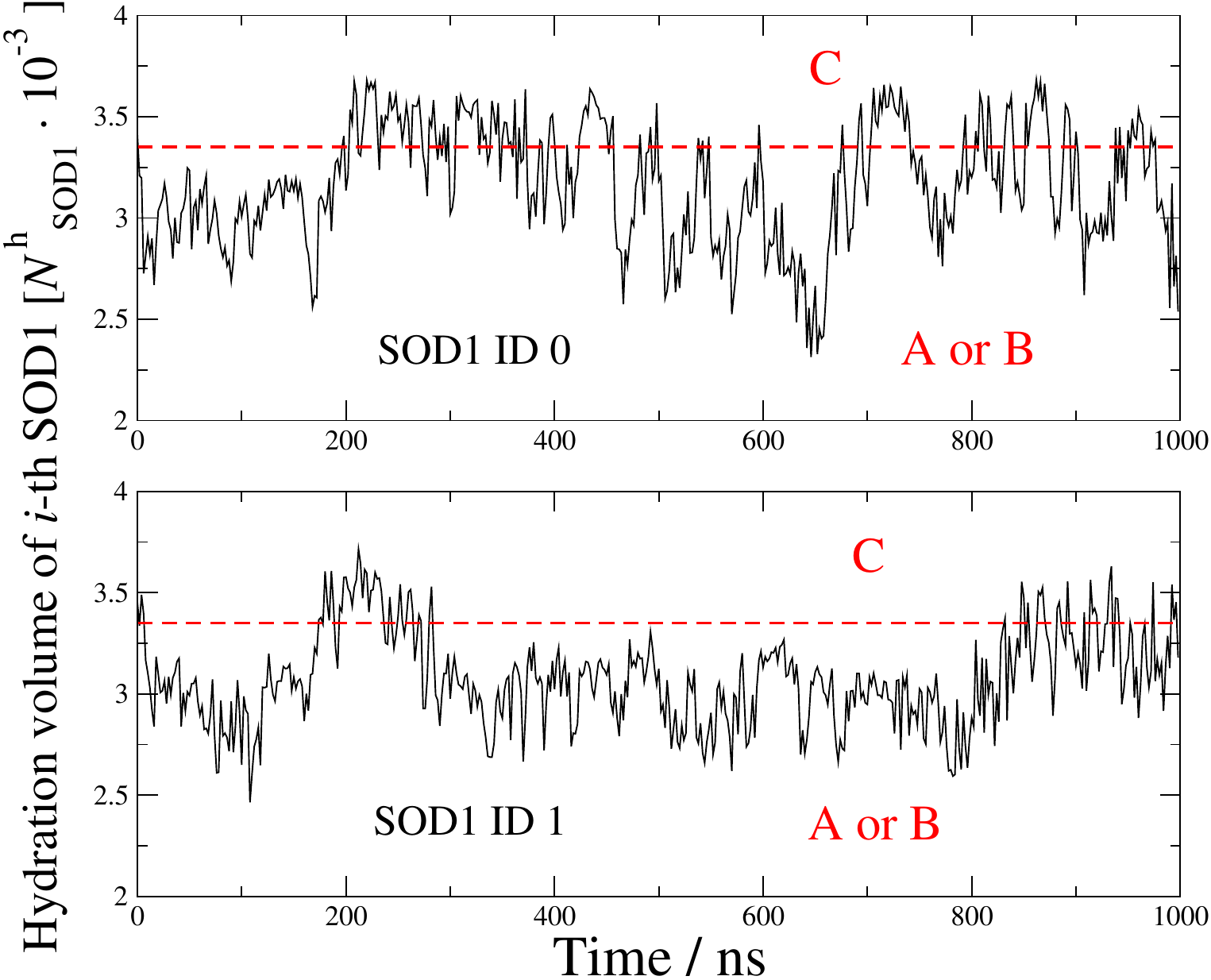}
    \includegraphics[width=0.28\textwidth]{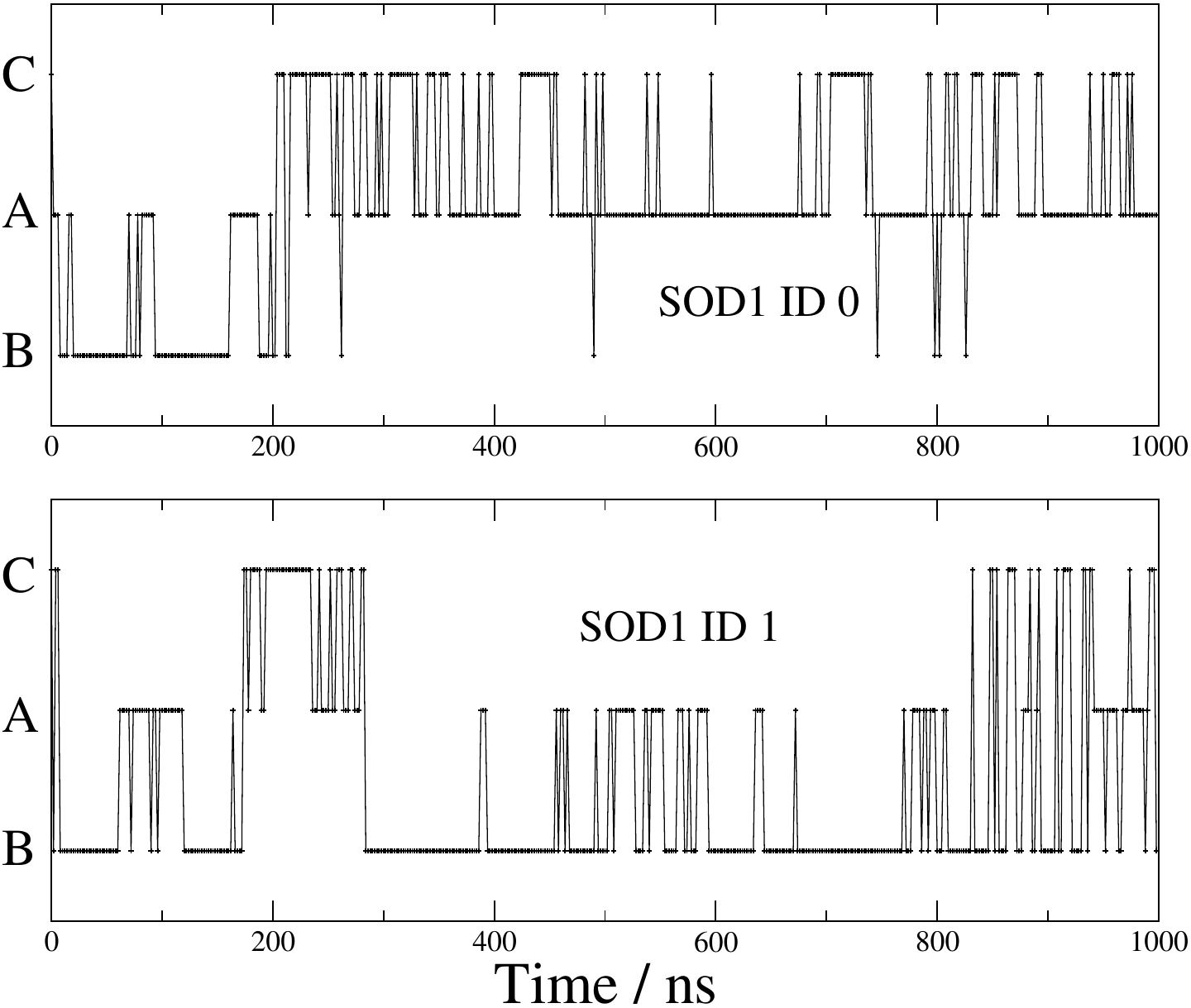}
    \includegraphics[width=0.28\textwidth]{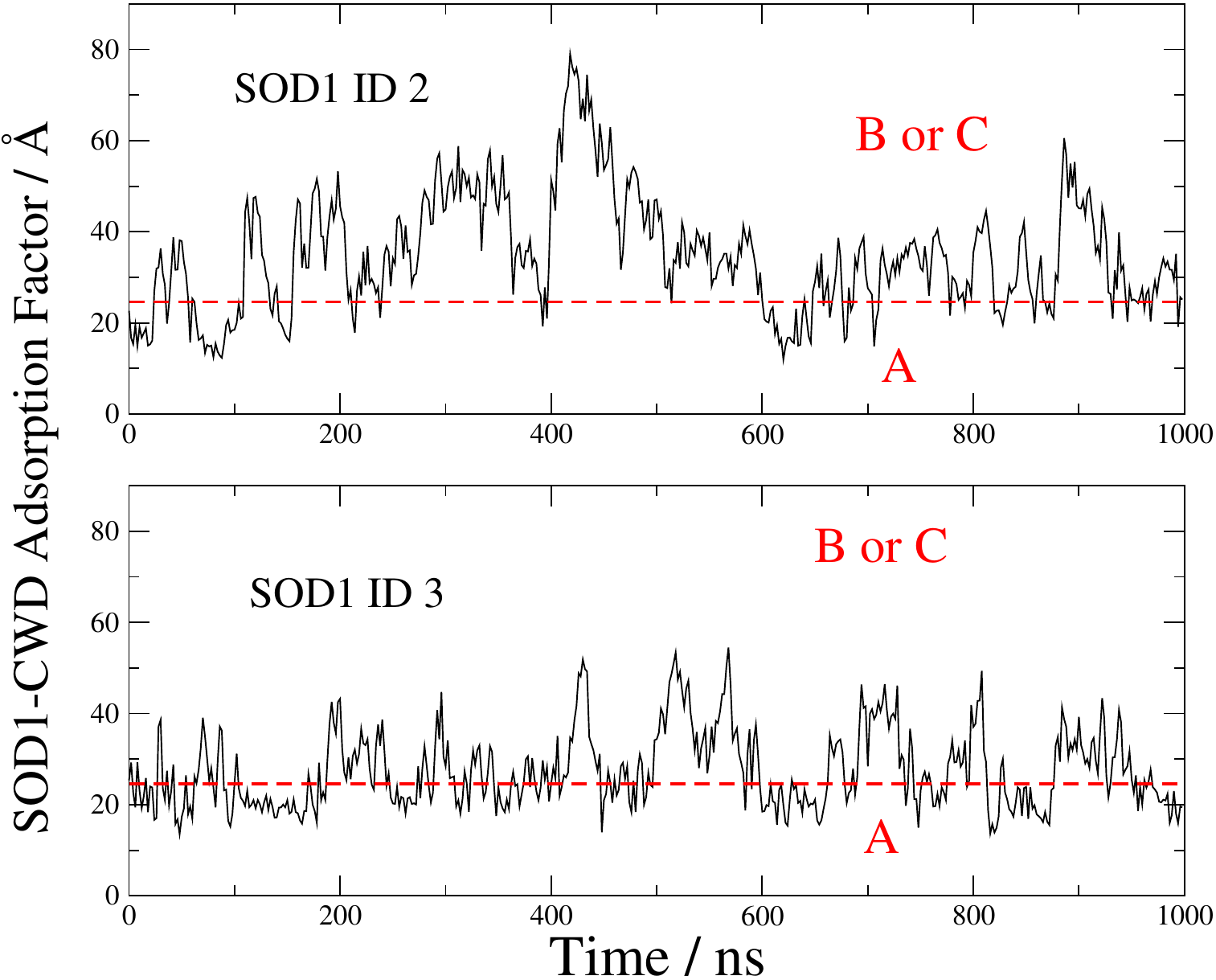} 
    \includegraphics[width=0.28\textwidth]{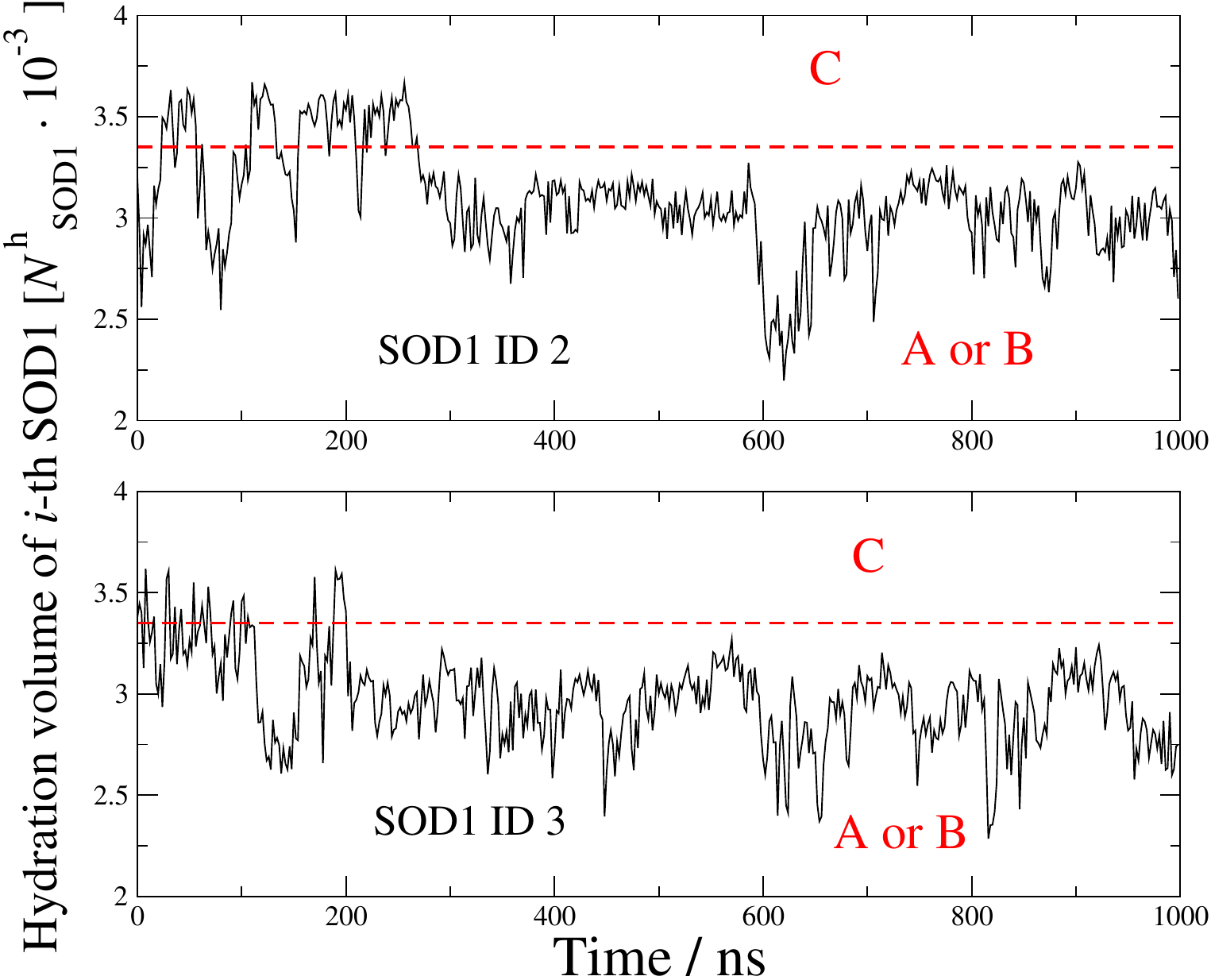}  
    \includegraphics[width=0.28\textwidth]{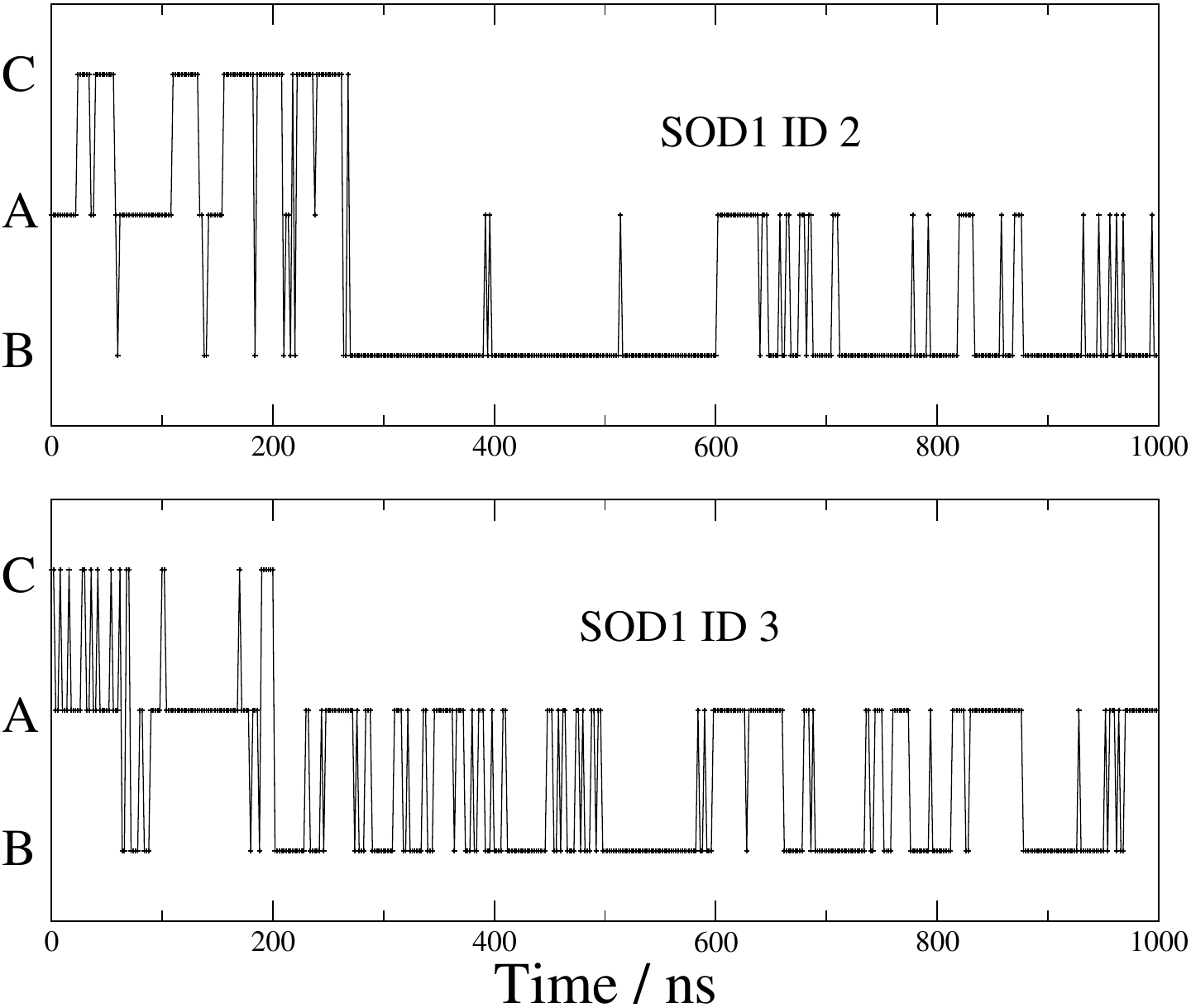}  
    \includegraphics[width=0.28\textwidth]{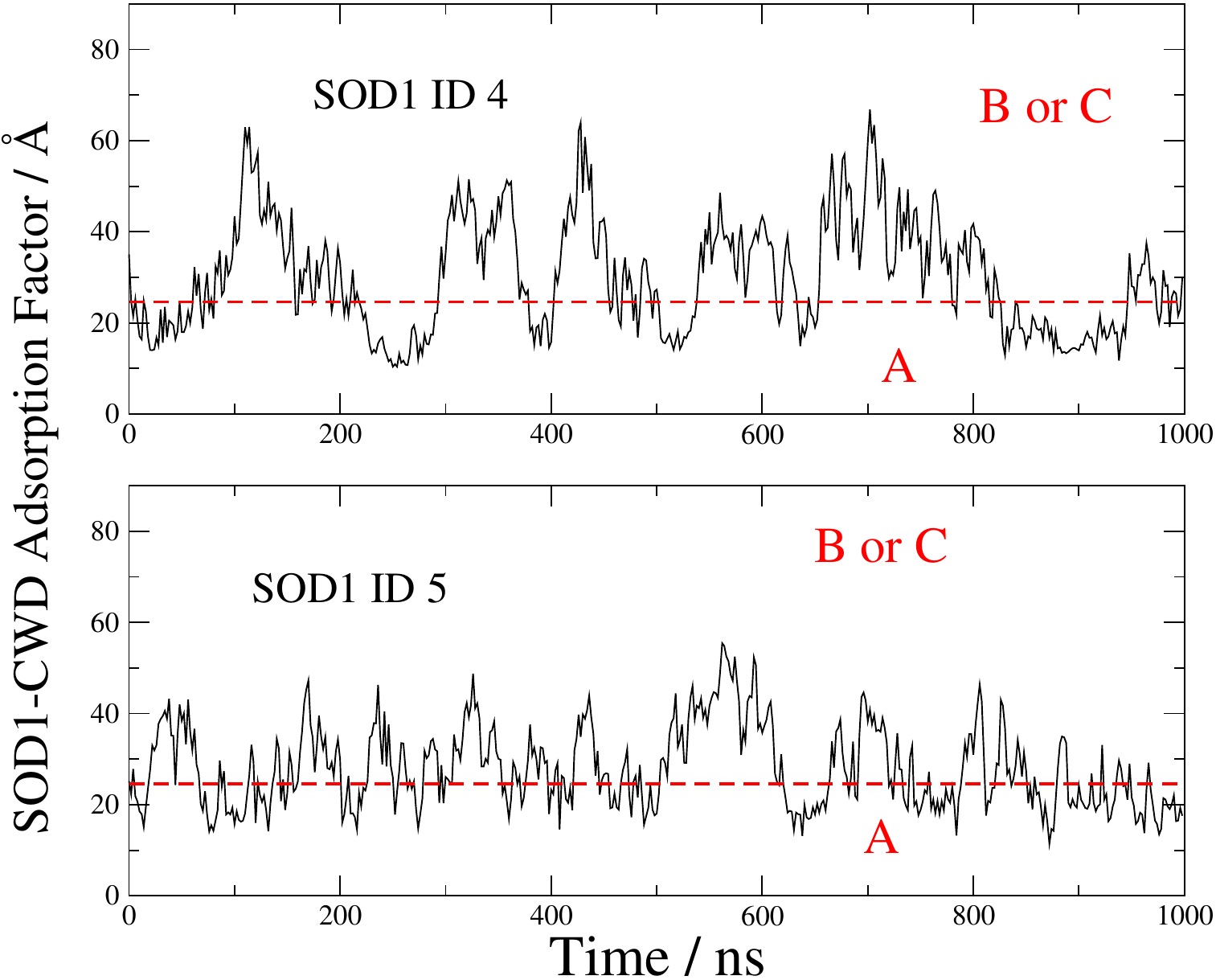} 
    \includegraphics[width=0.28\textwidth]{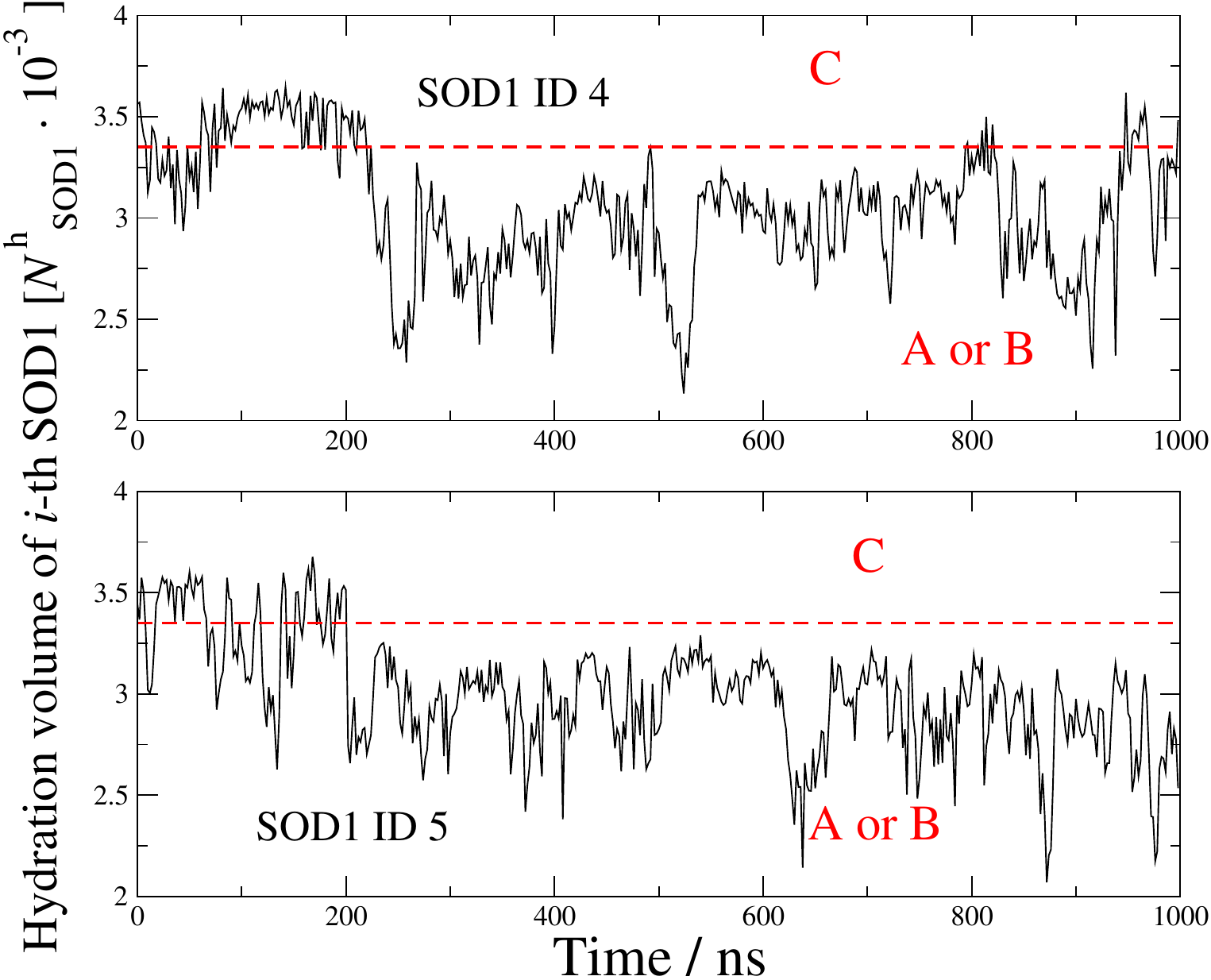} 
    \includegraphics[width=0.28\textwidth]{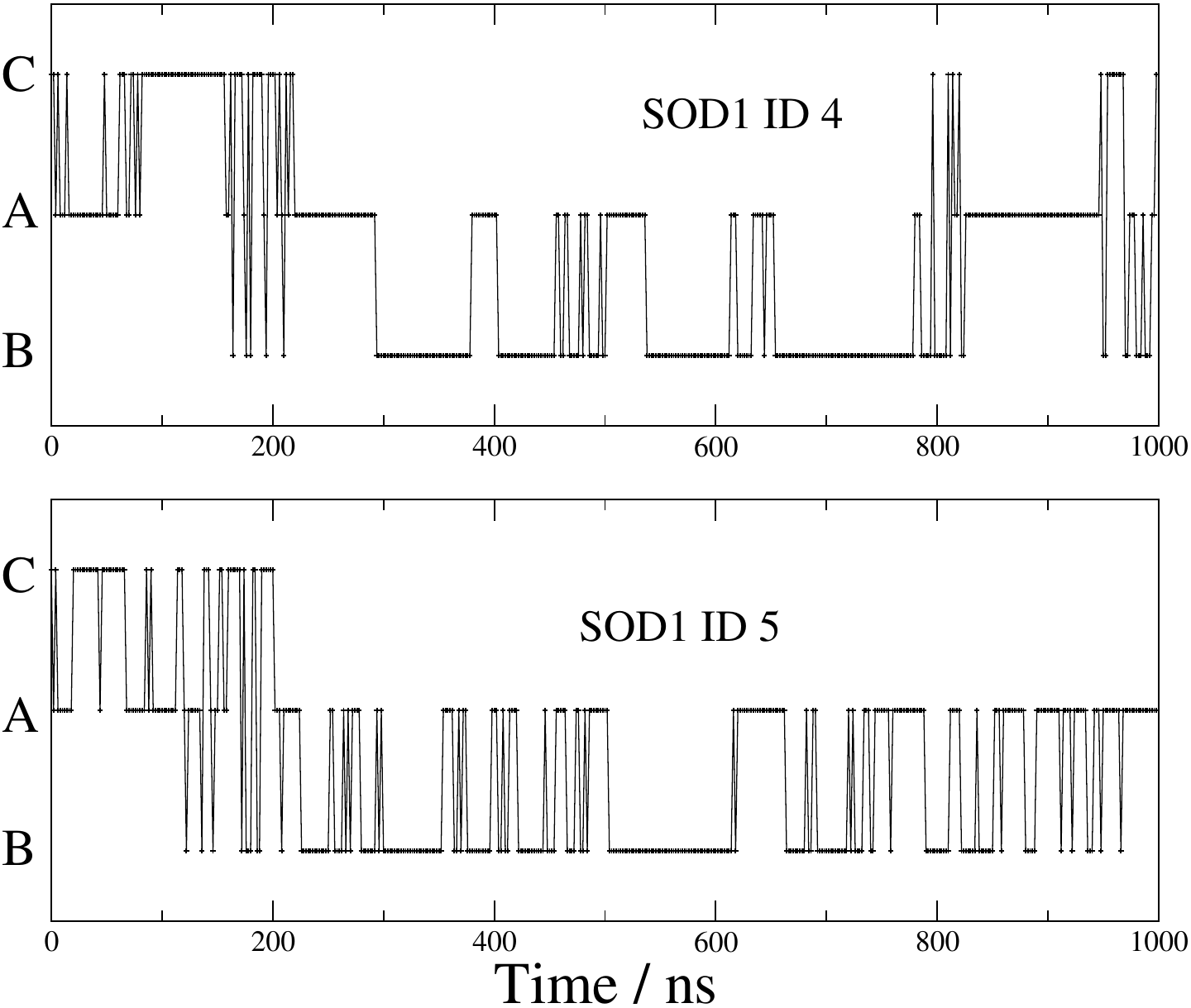} 
    \includegraphics[width=0.28\textwidth]{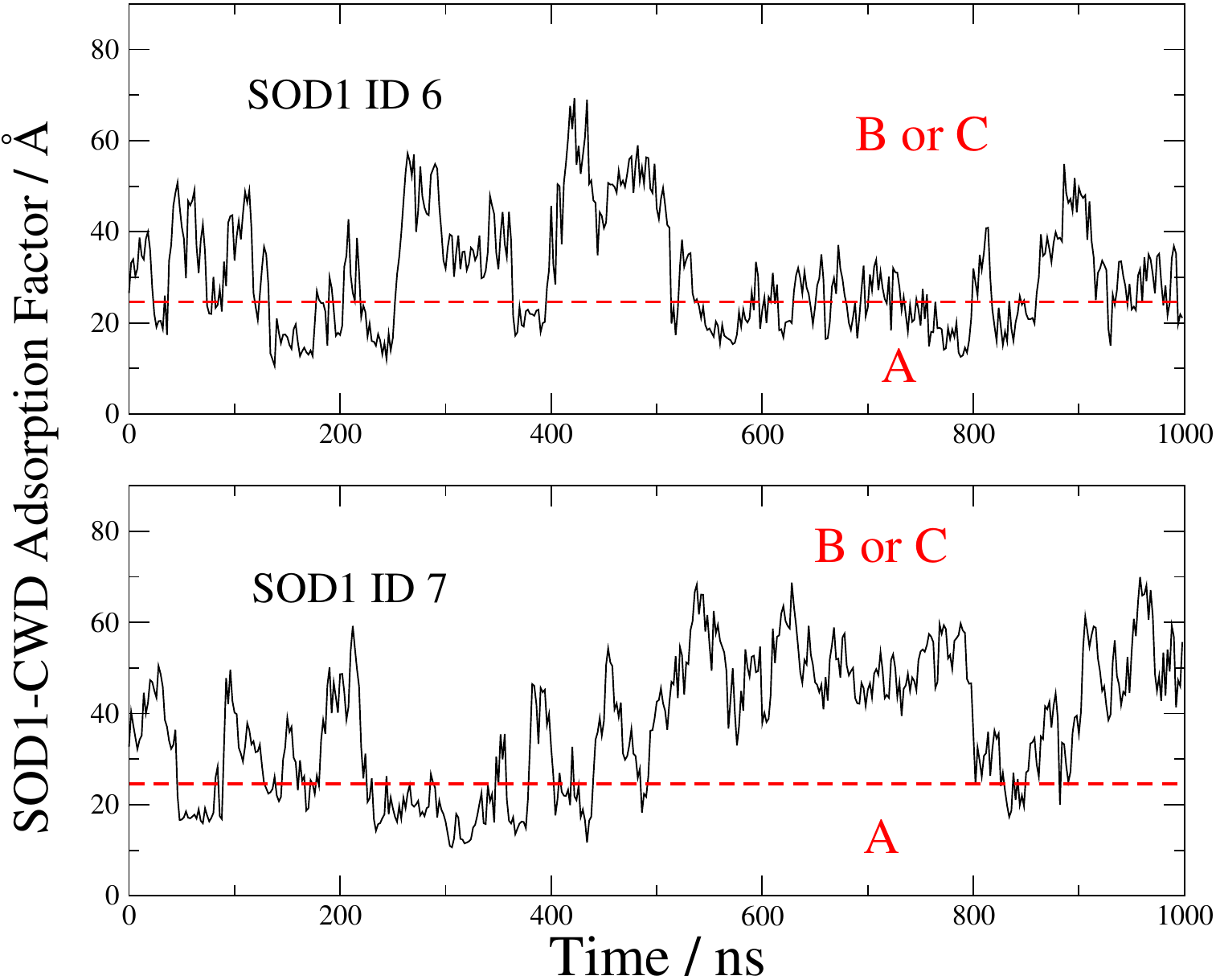} 
    \includegraphics[width=0.28\textwidth]{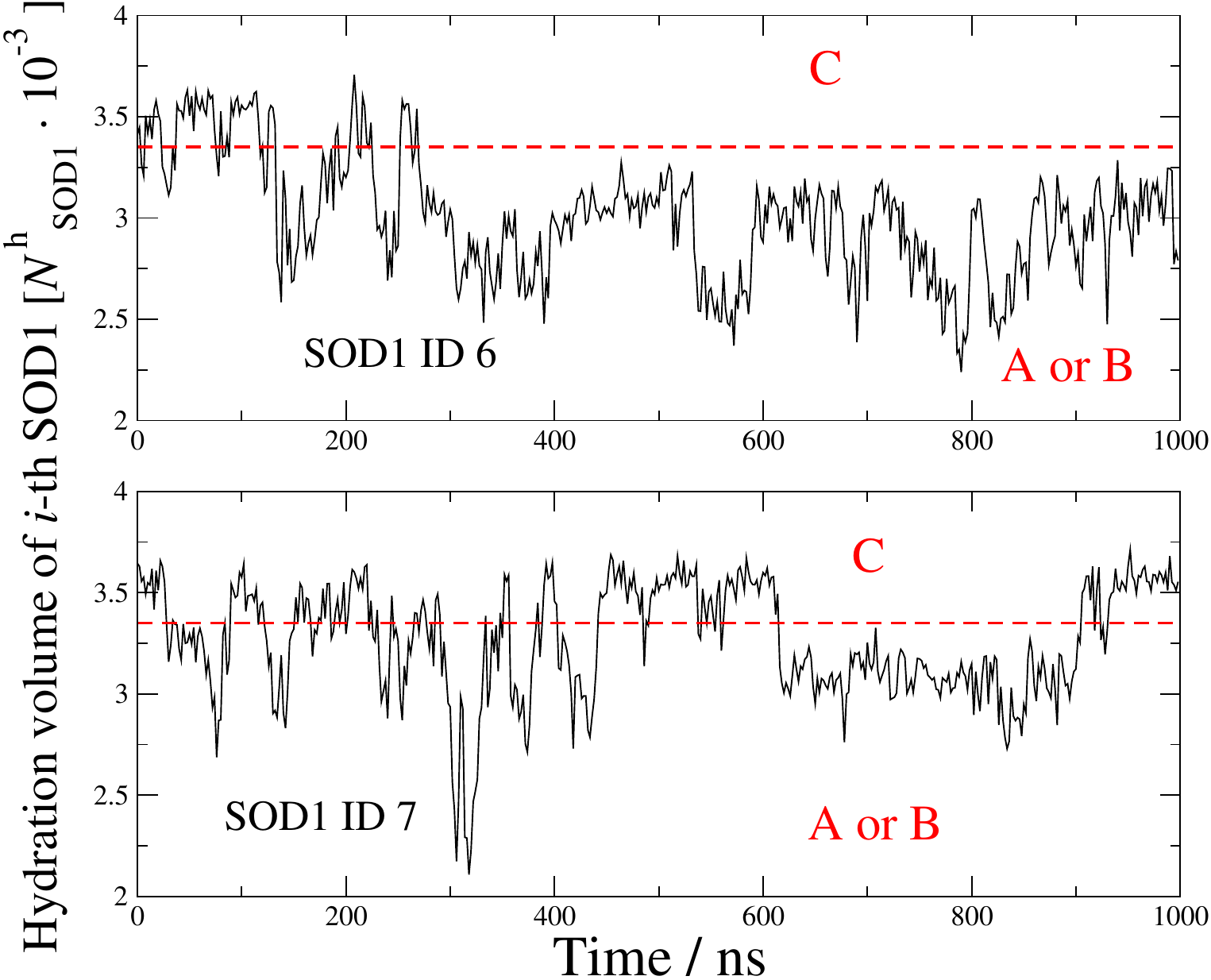}  
    \includegraphics[width=0.28\textwidth]{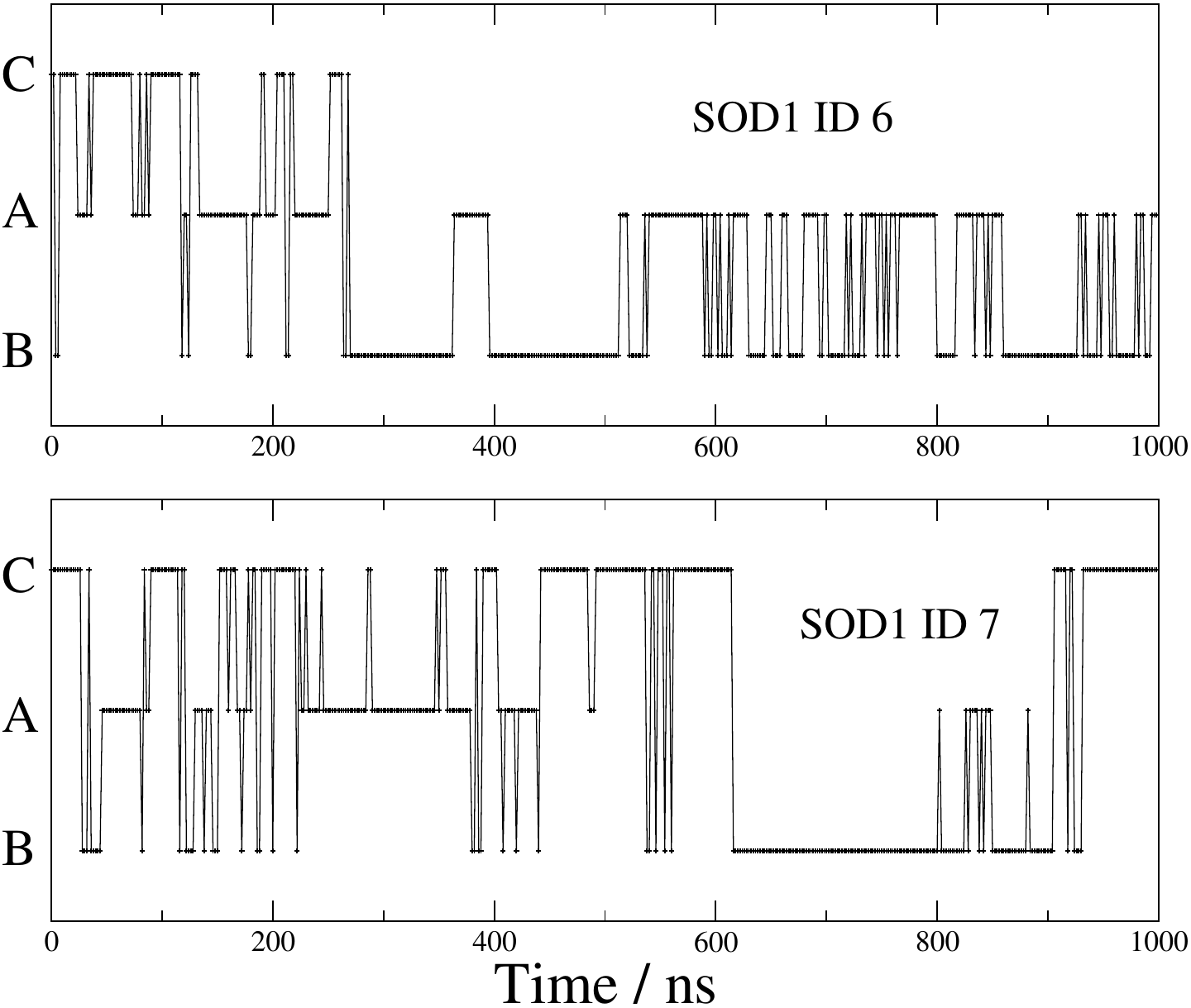} 
    \includegraphics[width=0.28\textwidth]{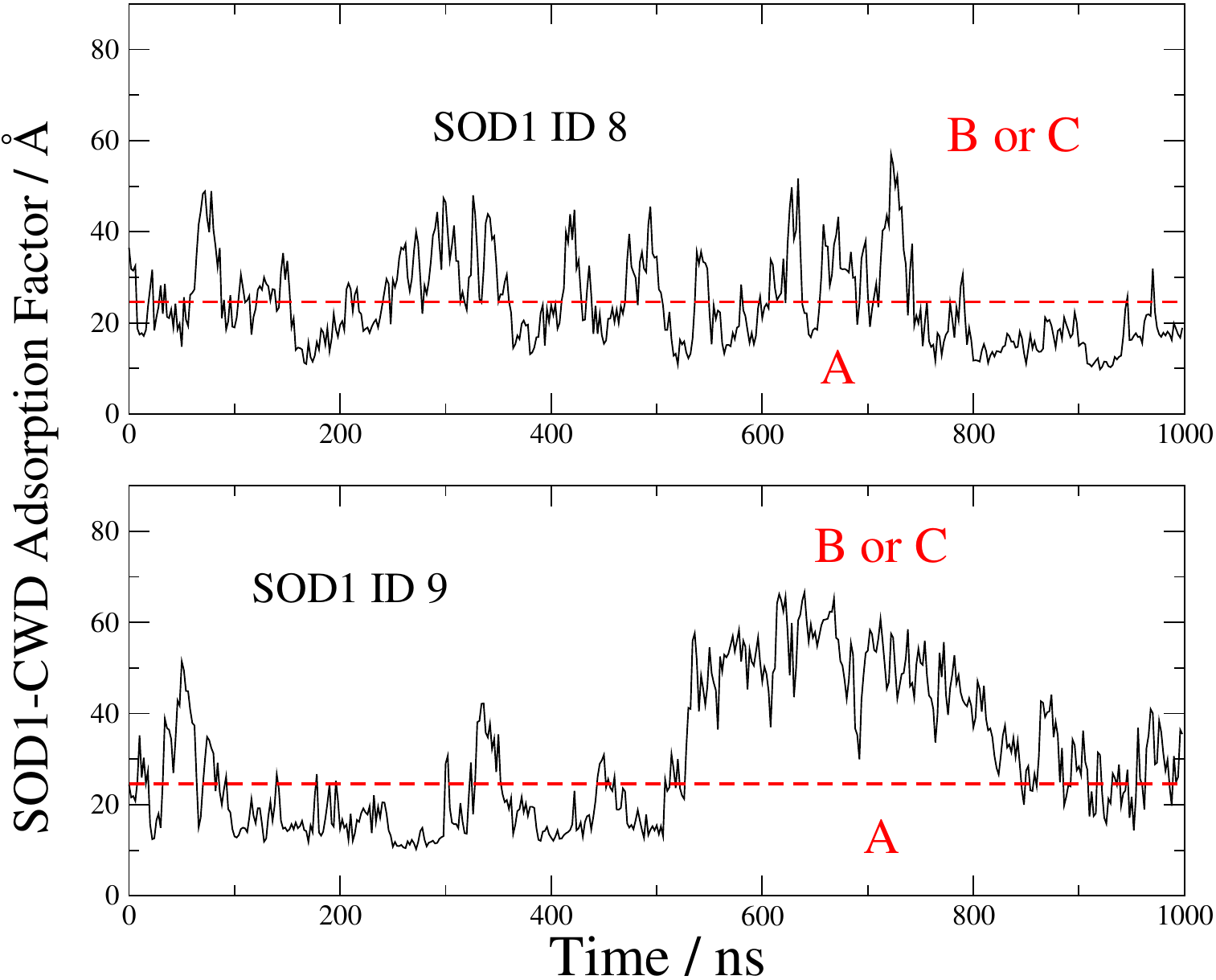} 
    \includegraphics[width=0.28\textwidth]{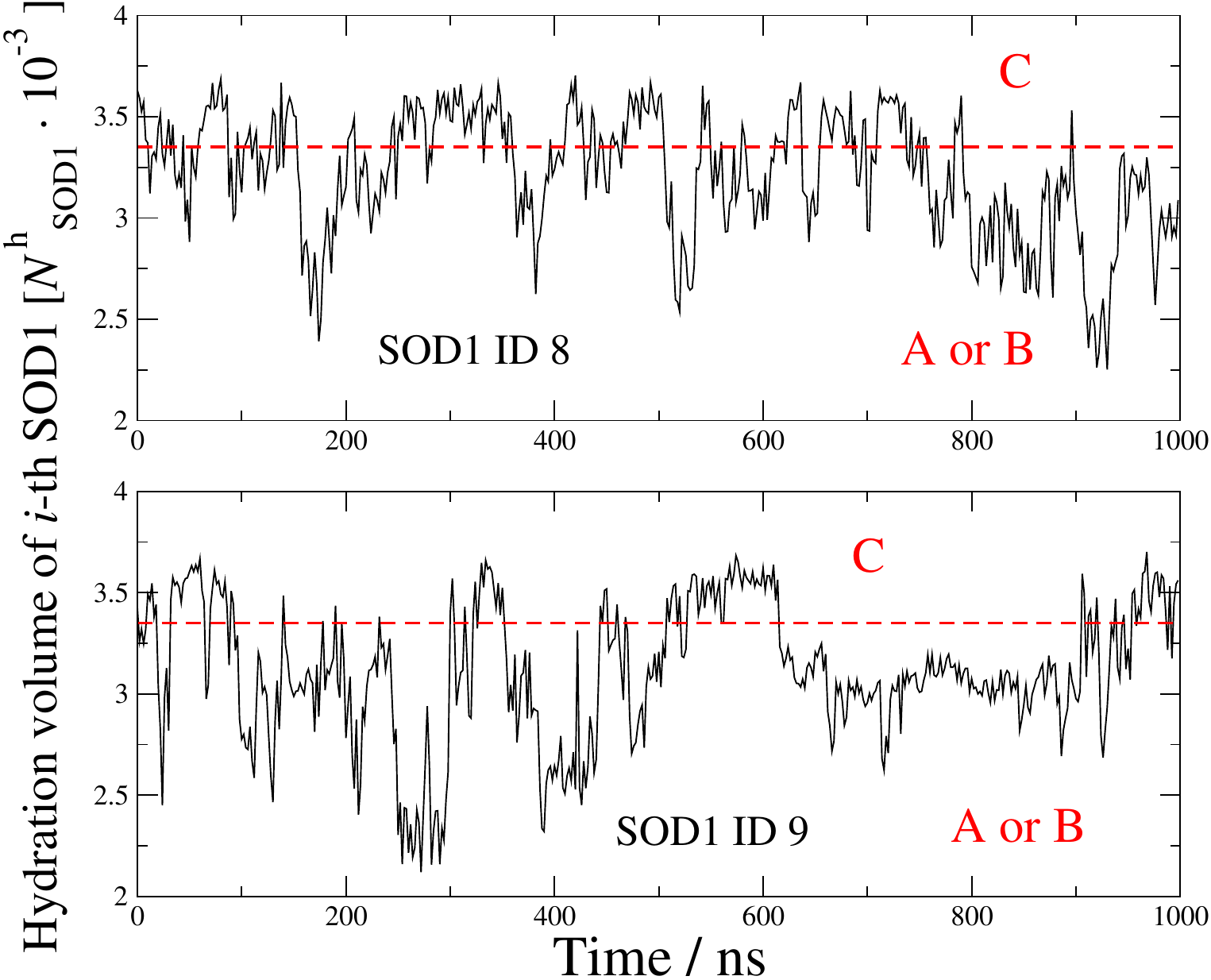} 
    \includegraphics[width=0.28\textwidth]{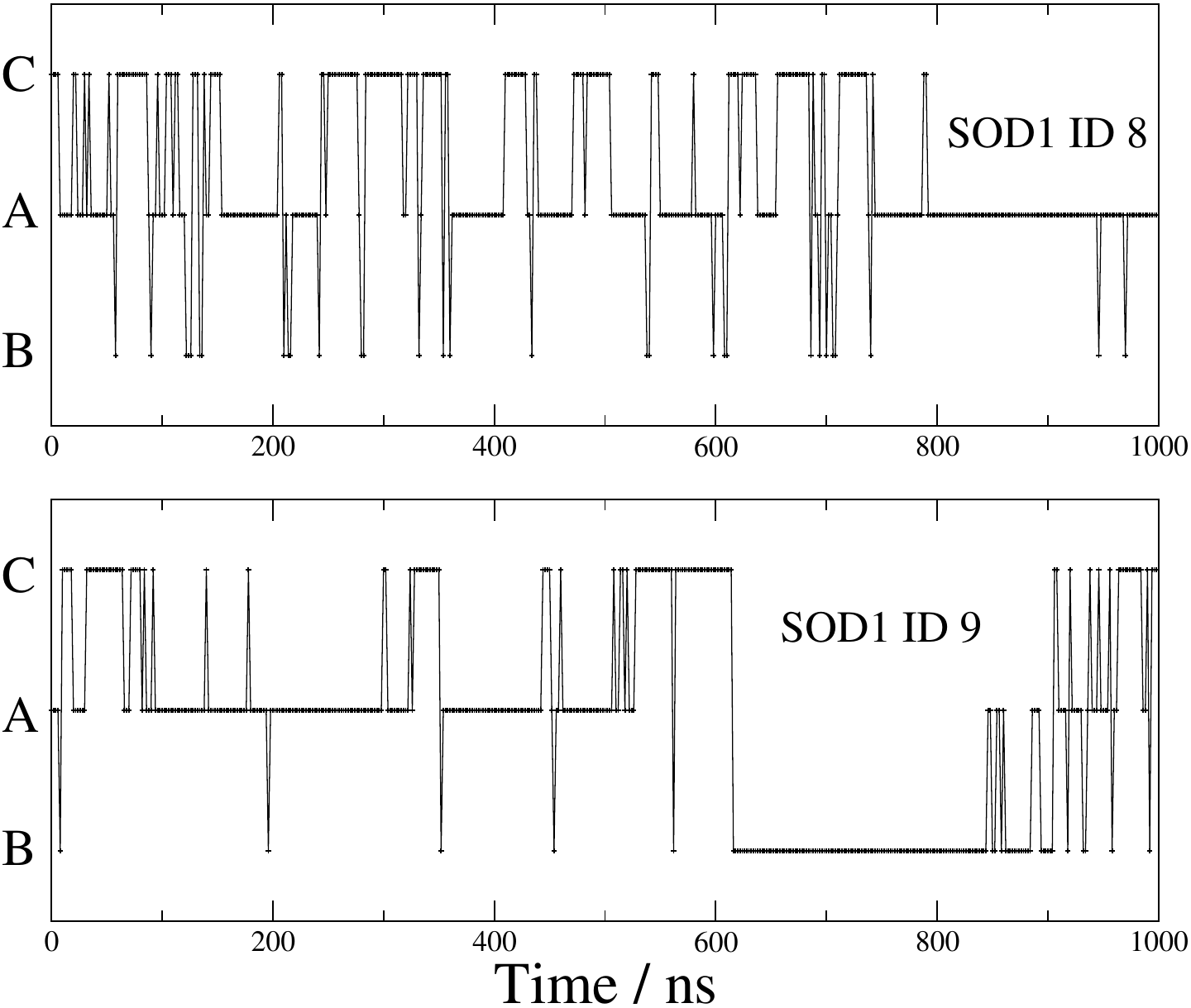} \\
    TIME $t [$~ns$]$ \\
    \caption{Time evolution of individual SOD1 in BSA solution ($x$-axis). From top to bottom, each of the SOD1s (ID from 0 to 9). The $y$-axis shows (left) the adsorption factor ${\cal C}_i$ in~\AA, (center) the number of water molecules that hydrate the SOD1 $N_{{\rm hyd, SOD1},i}\cdot 10^-3$, and (right) whether the SOD1 is in the A, B, or C state, as indicated at the top of the figure.}
    \label{fig:sod1_XY_timeseries}
\end{figure*}

\end{document}